%% file: main.tex
\documentclass[manuscript]{acmart}

\AtBeginDocument{%
  }

\setcopyright{none}
\renewcommand\footnotetextcopyrightpermission[1]{}

\title{``ChatGPT, what am I missing?'': Designing AI Workflows around Professional Task Structure to Shape Analytic AI Use}

\author{Zilin Ma}
\email{zilinma@g.harvard.edu}
\affiliation{%
  \institution{AI Institute, Harvard Business School}
  \city{Boston}
  \state{Massachusetts}
  \country{United States}}

\author{Suzi Jazmati}
\email{suzi.jazmati@frontline-associates.org}
\affiliation{%
  \institution{Frontline Associates}
  \city{Magog}
  \state{Quebec}
  \country{Canada}}

\author{Marco Chimenton}
\affiliation{%
  \institution{Independent Researcher}
  \city{Venice}
  \country{Italy}}

\author{Yiyang Mei}
\affiliation{%
  \institution{Emory University School of Law}
  \city{Atlanta}
  \state{Georgia}
  \country{United States}}

\author{Jacqueline Lane}
\affiliation{%
  \institution{Laboratory for Innovation Science at Harvard}
  \city{Cambridge}
  \state{Massachusetts}
  \country{United States}}

\author{Krzysztof Z. Gajos}
\affiliation{%
  \institution{Harvard John A. Paulson School of Engineering and Applied Sciences}
  \city{Allston}
  \state{Massachusetts}
  \country{United States}}

\author{Finale Doshi-Velez}
\affiliation{%
  \institution{Harvard John A. Paulson School of Engineering and Applied Sciences}
  \city{Cambridge}
  \state{Massachusetts}
  \country{United States}}

\begin{abstract}

General-purpose AI lets users choose what support to request, but leaves them to structure the support a professional task requires. We examine how interactive workflows can embed professional task structure without prescribing how users engage with AI. We designed two scaffolded interfaces around the same negotiation scaffold: one presented a completed AI analysis, while the other supported user-directed, incremental development. A four-condition randomized experiment with $N=\Nfinal$ participants compared these interfaces with no-AI and an AI chat interface. AI-supported conditions improved preparation coverage over unaided work; the scaffolded workflows further improved coverage over chat. Although the scaffolded workflows produced similar coverage, the user-directed workflow elicited a broader repertoire of analytic requests and lower subjective effort. Professional scaffolding therefore depends not only on displayed structure but on how workflows organize users’ engagement with it. Effective professional AI must structure how users and AI build analysis together.

\end{abstract}

\ccsdesc[500]{Human-centered computing~Human computer interaction (HCI)}
\ccsdesc[300]{Human-centered computing~Empirical studies in HCI}

\keywords{human--AI interaction, generative AI, cognitive labor, sensemaking, negotiation, agency, psychological ownership}

\usepackage{xspace}
\usepackage{placeins}
\usepackage{longtable}
\usepackage{array}

\newcommand{\Reader}{\textsc{Reader}\xspace}
\newcommand{\Chatbot}{\textsc{Chatbot}\xspace}
\newcommand{\Prefilled}{\textsc{AI-Prefilled}\xspace}
\newcommand{\Coevolving}{\textsc{Co-Evolving}\xspace}

\newcommand{\Nrecruited}{1{,}712}

\newcommand{\Nfinal}{800}
\newcommand{\Nreader}{181}
\newcommand{\Nchatbot}{260}
\newcommand{\Nprefilled}{206}
\newcommand{\Ncoevolving}{153}

\newcommand{\Tprep}{15}          
\newcommand{\Tassess}{17}        
\newcommand{\Nfiles}{15}         

\newcommand{\Pay}{\$7}
\newcommand{\Payrate}{\$11}

\begin{document}
\maketitle

\input{sections/01-introduction}
\input{sections/02-related-work}
\input{sections/04-study}
\input{sections/05-results}
\input{sections/06-discussion}
\input{sections/07-limitations}
\input{sections/08-conclusion}

\bibliographystyle{ACM-Reference-Format}
\bibliography{references}

\input{sections/09-appendix}

\end{document}

%% file: sections/01-introduction.tex
\section{Introduction}
\label{sec:introduction}

People using general-purpose interfaces such as ChatGPT for knowledge work may struggle to determine how AI can best support the task. With each prompt, users must diagnose what is difficult~\cite{tankelevitch2024metacognitive}, assess what they and the model can each contribute~\cite{dellacqua2026jagged,fugener2022delegation}, formulate an appropriate request~\cite{subramonyam2024gulf,zamfirescu2023johnny}, and evaluate the result~\cite{tankelevitch2024metacognitive,bucinca2021trust}. Many users lack this metacognitive knowledge. They prompt opportunistically and require explicit training to articulate what support they need~\cite{fugener2022delegation,zamfirescu2023johnny,ma2025rope,subramonyam2024gulf}. Consider a user seeking an exercise recommendation. They might delegate the entire task and ask AI to generate a plan or first develop a plan themselves and ask AI to critique it. The chat interface allows both of these approaches---and more---but provides little guidance on which approach to choose, and thus users may fail to recognize the best course of action. 

In many earlier decision-support systems, designers determined in advance what support the system would provide. Designers specified available interactions, organized tasks into workflows, or used scaffolds to make relevant questions and intermediate analyses visible~\cite{silver1990directed,silver1991guidance,todd1999strategy,russell1993cost,horvitz1999mixed}. A scaffold can guide users through a sound process. An exercise-planning system might ask for goals, injuries, and available equipment before producing a plan, encoding what a competent trainer would ask. However, it can also prescribe how the work should proceed and limit users' ability to decide what support to seek. That flexibility matters because useful cognitive support depends on the task and situation. Different users may need different forms of support. Some may need help challenging an initial commitment~\cite{schwenk1994devils,rastogi2022deciding}, organizing complex trade-offs~\cite{russell1993cost,suh2023sensecape}, or revisiting conflicting evidence~\cite{gero2024sensemaking}. Which form is useful may depend on task difficulty~\cite{steyvers2023three,bucinca2021trust,bucinca2025contrastive}, user experience~\cite{bucinca2026adaptive}, and time pressure~\cite{cao2023timepressure}.

This tradeoff between guiding users toward effective forms of support and preserving their flexibility to seek support as needed presents a challenge for designing AI systems for knowledge work. How can AI systems provide the benefits of professional scaffolding while preserving users’ ability to direct how they engage with AI? Prior decision-support research distinguishes between systems that provide recommended analyses or conclusions~\cite{parasuraman2000model,mosier1996automation} and systems that guide the process through which users arrive at them~\cite{silver1991guidance,todd1999strategy}. Generative AI can instantiate either approach. A scaffold can structure what AI produces, yielding a completed analysis for the user to inspect, or it can structure the interaction itself, allowing users to decide which parts to pursue while AI develops the corresponding analysis incrementally. When the same professional scaffold structures AI’s output versus the process of working with AI, how do these approaches affect work quality, users’ engagement with AI, and their experience of the process? This design choice matters wherever users bring professional frameworks into interacting with an AI, whether comparing policy options, making medical decisions, or preparing for a negotiation. Understanding how different workflow instantiations support these processes can help designers improve task quality while managing the effort required to produce and use the resulting analysis.

We study this question in frontline humanitarian negotiation. Negotiators often prepare under time pressure for discussions that they, rather than the AI, must conduct and defend. The Centre of Competence on Humanitarian Negotiation (CCHN) provides a framework for this preparation~\cite{cchn}. It asks negotiators to identify where the parties agree and disagree (the \textit{Island of Agreements}), distinguish stated positions from underlying interests (the \textit{Iceberg}), and establish what their organization cannot concede and the minimum outcome that must be secured (\textit{red lines} and \textit{bottom lines}). Together, these components describe what a thorough preparation should cover~\cite{cchn}. Moreover, professional negotiators increasingly use general-purpose chat assistants to produce these components~\cite{ma2025chatgpt}.

We developed two AI interfaces that embed the same professional negotiation scaffold but differ in when and how it is populated with AI-generated analysis. In \Prefilled{}, AI populates the scaffold before the user begins working with it, providing a completed analysis to inspect and use. In \Coevolving{}, the scaffold begins empty and organizes an interaction in which users decide what to pursue and AI develops the corresponding analysis incrementally.

\textbf{Providing a completed analysis} (\Prefilled{}) may support broad coverage while shifting users' effort toward reading and integrating AI-generated material. An \textbf{incremental workflow} (\Coevolving{}) allows users to choose which questions to pursue and which analyses AI develops, but requires continued interaction and may leave parts of the scaffold incomplete. Because directing a process does not require authoring its content, the two workflows may produce different patterns of AI use and different experiences of the resulting work\footnote{We studied decision agency and psychological ownership. Decision agency refers to users' sense that they control how the work proceeds, whereas psychological ownership refers to their sense that the resulting work is their own.}~\cite{draxler2024ghostwriter,lee2026relying,chi2026unowned}. To situate these workflows, we compared them with two baselines. \Reader{} provided case documents and a notepad without AI support, while \Chatbot{} added an open-ended assistant---the primary way negotiators currently interact with AI~\cite{ma2025chatgpt}. Together, the four conditions support three nested comparisons: AI-supported versus unaided preparation, scaffolded workflows versus open-ended chat, and incremental versus completed scaffold development. We specified directional hypotheses for preparation quality and research questions for AI use and user experience.

\begin{description}
\item[H1:] Pooling \Chatbot{}, \Prefilled{}, and \Coevolving{}, participants using AI-supported interfaces will produce higher-quality negotiation preparation than participants using \Reader{}.

\item[H2:] Pooling \Prefilled{} and \Coevolving{}, participants using scaffolded AI workflows will produce higher-quality preparation than participants using \Chatbot{}.

\item[RQ1:] How do the three AI-supported interfaces affect whether participants use AI and how they use it?

\item[RQ2:] How do the four interfaces shape subjective effort, decision agency, and psychological ownership?
\end{description}

We tested these hypotheses and research questions in a four-condition, between-subjects randomized experiment with $N=\Nfinal$ Prolific participants varying in prior negotiation experience. Participants completed a timed preparation exercise using a \Nfiles-document case co-developed from the CCHN Field Manual with an experienced frontline negotiator~\cite{cchn}. After preparation, we removed access to the AI systems and case files but allowed participants to retain their transcripts and preparation artifacts while answering written questions. We evaluated their answers using negotiator-developed scoring criteria.

Pooling the two scaffolded workflows produced stronger preparation coverage than open-ended \Chatbot{}, although \Prefilled{} and \Coevolving{} did not differ detectably in overall coverage (Figure~\ref{fig:nested-outcome-comparisons}). The workflows nevertheless shaped how participants used AI. \Chatbot{} participants were more likely to send case passages without an accompanying request, and requests in both \Chatbot{} and \Prefilled{} centered on simplifying existing information. \Coevolving{}, by contrast, elicited a broader repertoire spanning own party, counterpart, cross-party, and strategic analysis, while requiring less subjective effort than \Prefilled{} (Figures~\ref{fig:prompt-use-patterns} and~\ref{fig:nested-outcome-comparisons}). The improvements in quality also followed what the scaffold represented: several dimensions of factual and package preparation improved, but package-risk coverage did not, and requests to verify or challenge the analysis remained uncommon (Figure~\ref{fig:prompt-use-patterns}; Appendix Figure~\ref{fig:question-quality-comparisons}).

These findings suggest three implications for designing AI-supported knowledge work:
\begin{itemize}

\item \textbf{Open-ended chatbots do not remove designers' responsibility to scaffold the work.} Although chat makes many forms of assistance available, it leaves users responsible for recognizing what matters, decomposing the task, and formulating productive requests. Professional task structure can reduce this burden by making relevant questions and intermediate analyses visible instead of requiring users to reconstruct a professional workflow through prompting.

\item \textbf{How a scaffold is instantiated shapes how users work.} \Prefilled{} and \Coevolving{} produced similar preparation coverage but different patterns of engagement and effort. Compared with receiving a completed analysis, directing its incremental development elicited a broader analytic repertoire and lower subjective effort (Figures~\ref{fig:prompt-use-patterns} and~\ref{fig:nested-outcome-comparisons}). Workflow design should therefore consider how users produce and experience an analysis alongside its quality, evaluating engagement, effort, agency, and ownership separately.


\item \textbf{Designing professional AI is a problem of work design.} Adding AI does not eliminate decisions about task structure, initiative, authorship, and verification. These decisions instead become embedded in how the workflow allocates work between users and AI. Effective professional AI should therefore organize not only what the system produces, but also how users direct, inspect, and develop that work.

\end{itemize}

%% file: sections/02-related-work.tex
\section{Related Work}
\label{sec:related-work}

\subsection{Configuring AI Decision support}

Decision-support interfaces determine both what a system does and what remains for the user. Automation can operate at different stages of work, including information acquisition, analysis, decision selection, and action, and each stage can be automated to different degrees~\cite{parasuraman2000model}. Decision-support research further distinguishes \textit{system restrictiveness}, which limits the strategies or sequences available to users, from \textit{decisional guidance}, which provides information, recommends a strategy, or helps execute one~\cite{silver1990directed,silver1991guidance}. Even when several strategies remain available, an interface can change their cognitive cost and thereby influence which strategy users adopt~\cite{todd1999strategy}. Mixed-initiative systems make a related design choice by determining when the user or system should take the initiative~\cite{horvitz1999mixed}. In each case, designers configure in advance how the system will participate in the work.

Open-ended generative AI shifts much of this choice to users by allowing queries through language. Without a single predefined workflow~\cite{retkowsky2023playmate}, users can ask a chat interface to retrieve information, summarize evidence, compare alternatives, develop an analysis, or challenge a conclusion. They therefore decide through individual prompts what to delegate and how AI should contribute~\cite{tankelevitch2024metacognitive,randazzo2026cyborgs}. These prompts allocate cognitive work between the user and the system rather than merely communicate instructions~\cite{baird2021delegation}.

This flexibility creates a metacognitive demand. Users must recognize what support the task requires, assess their own and the model's capabilities, formulate an executable request, and evaluate the response. People often lack the metaknowledge needed to allocate work productively~\cite{fugener2022delegation}. They may explore prompts opportunistically, struggle to translate an intention into a request, or follow a recommendation without recognizing when it is wrong~\cite{zamfirescu2023johnny,subramonyam2024gulf,khurana2024whyandwhen}. Training users to articulate requirements can improve outputs, but conventional prompt-engineering instruction does not necessarily close this gap~\cite{ma2025rope}. AI literacy similarly treats effective use of generative AI as a set of individual capabilities for understanding and evaluating AI~\cite{long2020ailiteracy}. These approaches help users operate an open-ended system, but they leave each user responsible for constructing the workflow through which AI supports the task.

Adaptive and generative interfaces shift some of this burden back into the system. Adaptive systems learn which assistance to provide to a particular user~\cite{bucinca2026adaptive}. Generative interfaces construct tools or controls around a stated objective, while just-in-time systems infer an objective from users' activity and generate support for it~\cite{leviathan2026generativeui,lam2026justintime,vaithilingam2024dynavis,cao2025malleable}. These approaches reduce the need to select a fixed tool, but they still depend on an objective expressed by the user or inferred by the system. They do not necessarily make visible the larger professional task of which that objective is one part.

\subsection{Professional scaffolds as output representations and workflow designs}

A professional scaffold externalizes recurring components of competent work. It can identify questions that should be considered, intermediate analyses that should be developed, and relationships that should be examined. External representations have long supported sensemaking by helping people organize evidence, retain intermediate results, and revise their understanding of a problem~\cite{russell1993cost}. LLM-based systems extend this approach by organizing literature, representing design spaces, and organizing collaborative ideas~\cite{kang2023synergi,suh2023sensecape,suh2024luminate,he2024ai}. Other systems provide organized overviews, diagrams, and comparisons that help users inspect relationships among model outputs~\cite{gero2024sensemaking,jiang2023graphologue,arawjo2024chainforge}. Such representations reduce the need for users to reconstruct the professional scaffold from memory or through a series of independent prompts.

The same scaffold can play different roles in human--AI work. A system can perform the analysis and present the result within a professional scaffold, automating part of information analysis while leaving the user to inspect and apply it~\cite{parasuraman2000model,mosier1996automation}. Alternatively, the scaffold can organize the process through which the analysis is produced. Its components can become targets for interaction, allowing users to choose what to examine while AI develops the corresponding analysis. This resembles decisional guidance that helps users organize and execute a process rather than simply delivering its result~\cite{silver1991guidance,todd1999strategy}. The first approach uses the scaffold primarily to organize AI's output. The second also uses it to organize how the user and AI work together.

Existing studies have compared scaffolded and open-ended AI interfaces, but they have not compared different ways of using the same scaffold. For example, guidance around an AI course assistant reduced unfocused queries and assignment-copying requests~\cite{kumar2025math}, while a scaffolded prompting platform elicited behaviors associated with learning gains, although these behaviors did not improve performance or persist in an unconstrained interface~\cite{brender2025structured}. Both studies changed the scaffold presented to users and the process users followed at the same time.

Scaffolds can also direct attention selectively and create blind spots. A representation makes the components it names easier to notice, but cannot represent every consideration that competent work may require. Decision aids can produce an \textit{unprompted-item blind spot}, in which users attend to considerations named by the aid and overlook relevant considerations it omits~\cite{seow2011restrictiveness}. A professional scaffold may therefore improve the parts of a task it represents without improving omitted components. Making a scaffold visible also does not ensure that users will verify AI-generated content or recognize when the scaffold itself is incomplete. Prior work therefore does not establish how presenting the same professional scaffold as a completed AI analysis rather than a user-directed workflow changes interaction, or whether either implementation improves parts of the task that the scaffold omits.

\subsection{Consequences for engagement and work experience}

In addition to work quality, how a scaffold allocates analytic work may affect both engagement and the experience of work. For example, AI assistance can reduce independent reasoning or unassisted performance~\cite{bastani2025generative,gajos2022people,lee2025impactgenai}. People may substitute automated judgments for their own vigilance, place excessive weight on algorithmic advice, or follow incorrect suggestions~\cite{mosier1996automation,logg2019algorithm,spatharioti2025search}. 

AI can also redistribute effort rather than simply reduce it. Generative AI may move work from producing an answer to formulating a request, evaluating the response, and integrating it into a larger task~\cite{tankelevitch2024metacognitive}. Receiving a completed analysis may reduce the effort of producing content while creating work in reading, interpreting, and integrating a large body of AI-generated material. Developing the analysis incrementally requires continued interaction, but may let users control what is generated and process it in smaller parts. Prior research does not establish which arrangement will feel less effortful when both use the same professional scaffold.

Decision agency and psychological ownership capture different experiences of this allocation. Decision agency concerns whether users experience themselves as controlling how the work proceeds, including what to pursue, accept, revise, or ignore. Psychological ownership concerns whether they experience the resulting work as their own. Control, intimate knowledge, and self-investment provide separate routes to psychological ownership~\cite{pierce2001ownership}. A person may therefore direct an AI-supported process without experiencing authorship of the content AI produces~\cite{draxler2024ghostwriter}. Passive use of AI-generated content has been associated with lower self-efficacy, ownership, and meaning than drafting content before asking AI to refine it~\cite{lee2026relying}. AI-generated goals can likewise improve immediate quality while reducing ownership, commitment, and follow-through~\cite{chi2026unowned}. 

\subsection{AI support for negotiation preparation}

Most AI negotiation research treats AI as the negotiator. Automated agents bargain on behalf of a principal under explicit preference models~\cite{lopes2008automated}, language agents are evaluated against one another in scorable bargaining games~\cite{abdelnabi2024stakeholders}, and negotiation competitions examine which agent behaviors predict agreement outcomes~\cite{vaccaro2026pnas}. This literature asks whether people should delegate negotiation to an autonomous system or remain active participants in it~\cite{gratch2026agents,curhan2026competitions}.

Frontline humanitarian negotiation presents a different design problem. Negotiators prepare for conversations that they must conduct, revise, and defend themselves. Practitioner doctrine provides a professional scaffold for this preparation by specifying recurring analyses that negotiators should develop~\cite{cchn}. Field-informed research finds that practitioners value AI for retrieving and organizing information but resist systems that turn contextual judgments into prescriptive recommendations~\cite{ma2025chatgpt}. This setting therefore requires both scaffolding and flexibility. The interface can make the components of competent preparation visible, but negotiators must retain the ability to decide which issues require attention and how AI should support them.

%% file: sections/04-study.tex
\section{Method}
\label{sec:method}

\subsection{Interface overview}

\begin{figure*}[t]
  \centering
  \begin{minipage}[t]{0.485\textwidth}
    \vspace{0pt}
    \includegraphics[width=\linewidth]{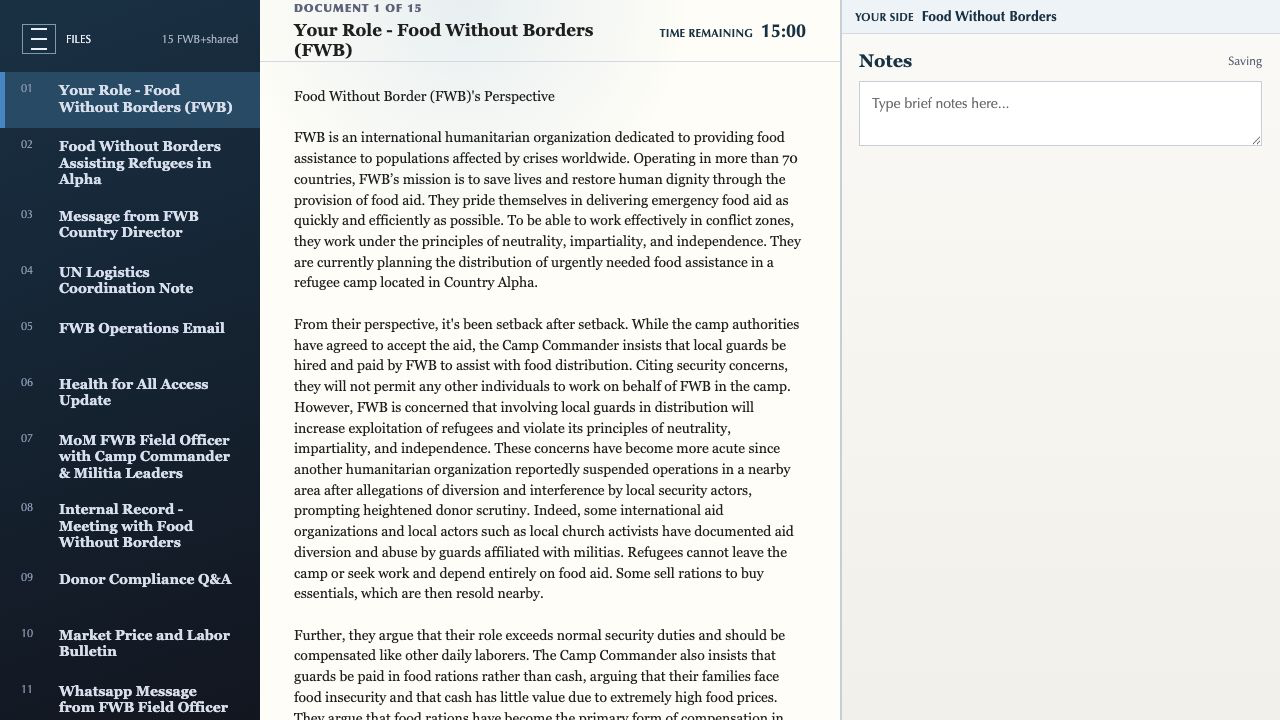}
    \centering\small\textbf{(a) \Reader}
  \end{minipage}\hfill
  \begin{minipage}[t]{0.485\textwidth}
    \vspace{0pt}
    \includegraphics[width=\linewidth]{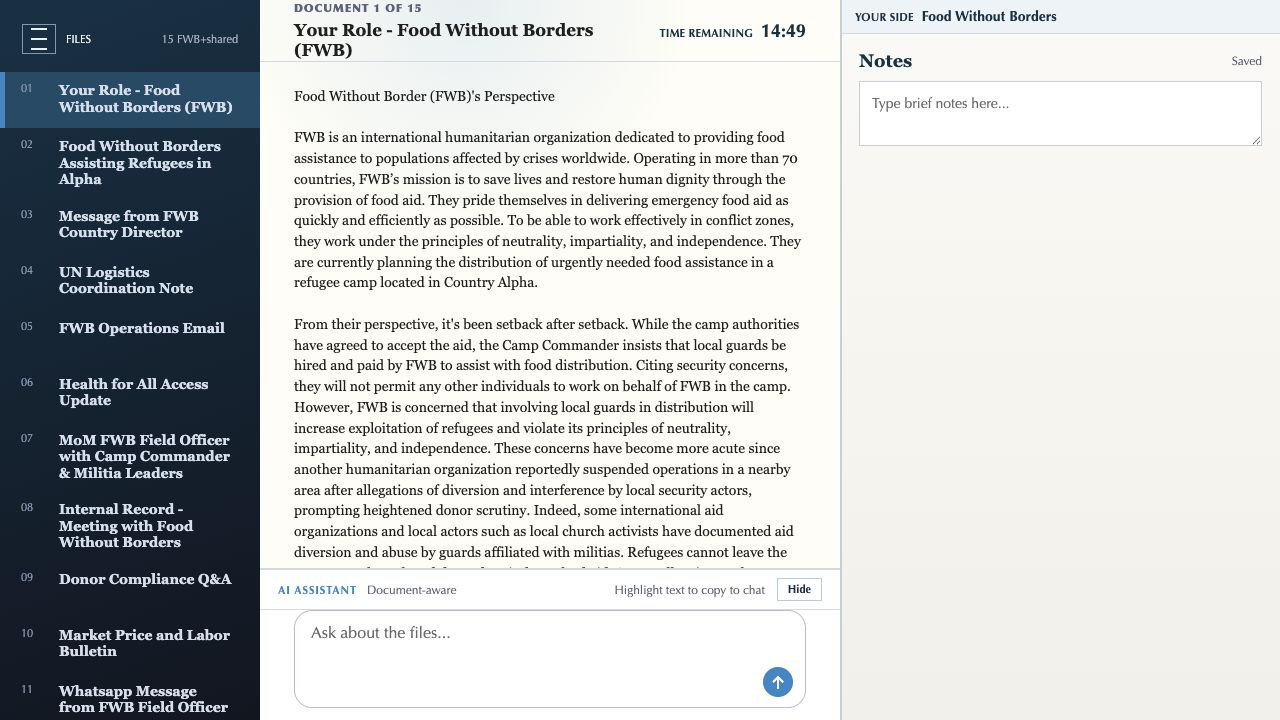}
    \centering\small\textbf{(b) \Chatbot}
  \end{minipage}\hfill
  \vspace{0.6em}

  \begin{minipage}[t]{0.485\textwidth}
    \vspace{0pt}
    \includegraphics[width=\linewidth]{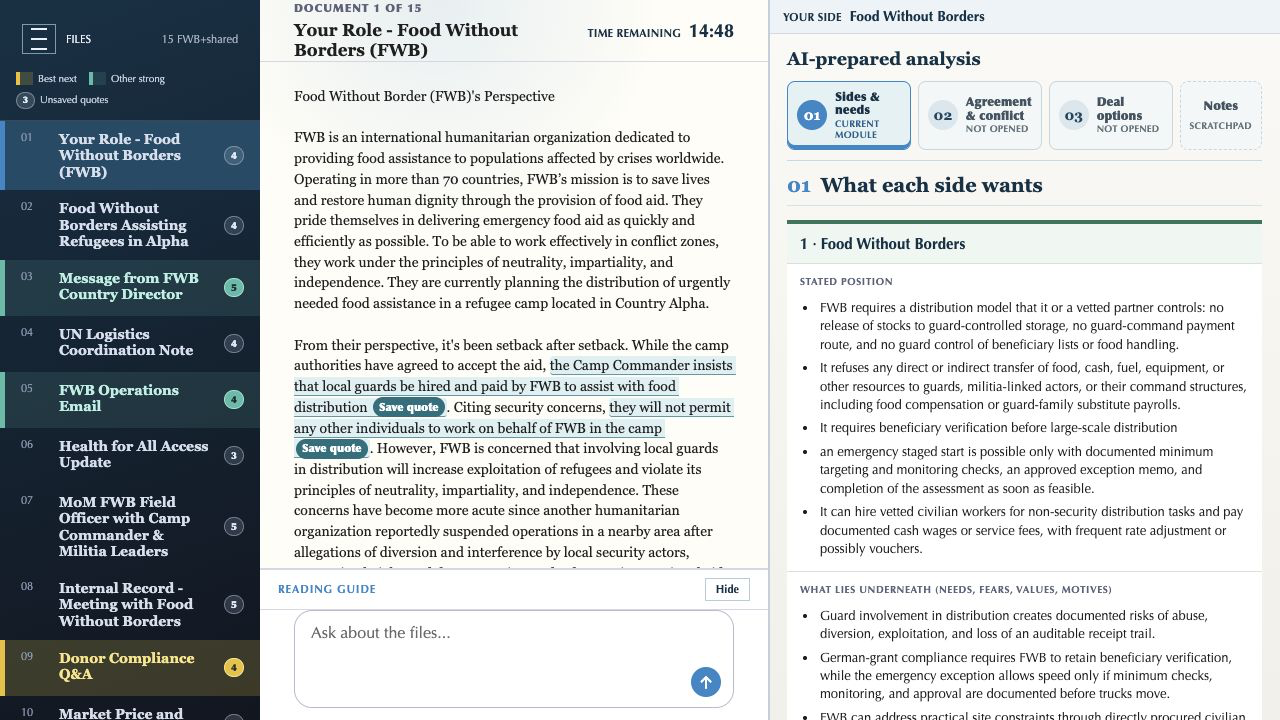}
    \centering\small\textbf{(c) \Prefilled}
  \end{minipage}\hfill
  \begin{minipage}[t]{0.485\textwidth}
    \vspace{0pt}
    \includegraphics[width=\linewidth]{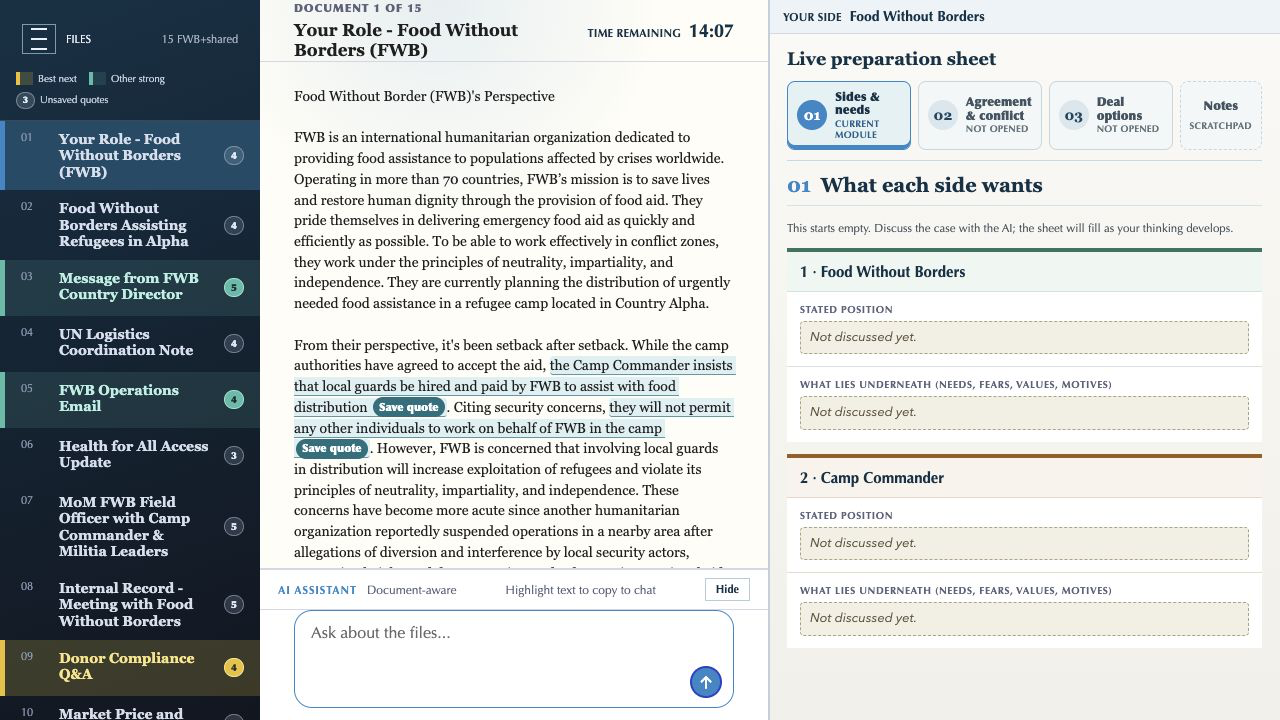}
    \centering\small\textbf{(d) \Coevolving}
  \end{minipage}

  \vspace{0.6em}

  \begin{minipage}[t]{0.20\textwidth}
    \vspace{0pt}
    \includegraphics[width=\linewidth,trim=0 470 0 0,clip]{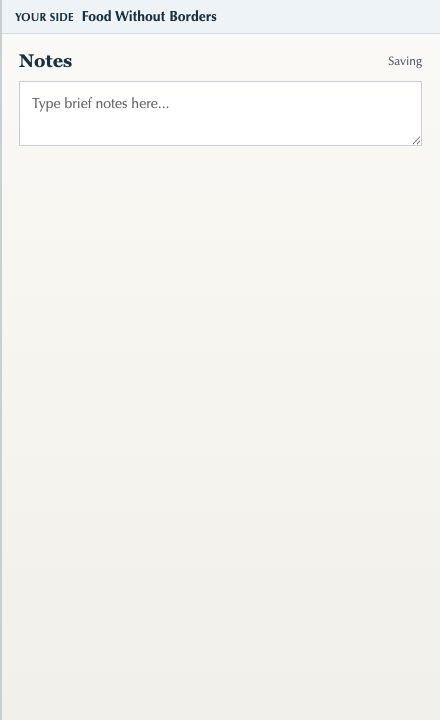}
    \centering\footnotesize\textbf{(e) \Reader}\\private notes
  \end{minipage}\hfill
  \begin{minipage}[t]{0.28\textwidth}
    \vspace{0pt}
    \includegraphics[width=\linewidth]{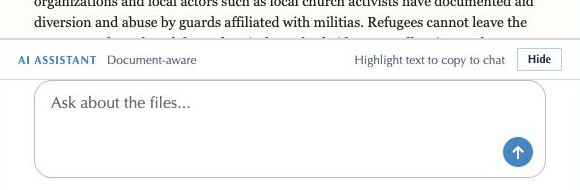}
    \centering\footnotesize\textbf{(f) \Chatbot}\\document-aware assistant
  \end{minipage}\hfill
  \begin{minipage}[t]{0.22\textwidth}
    \vspace{0pt}
    \includegraphics[width=\linewidth,trim=0 270 0 0,clip]{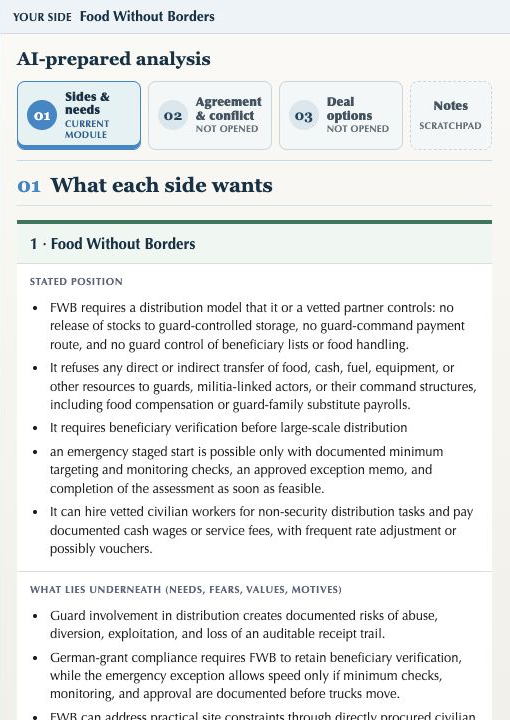}
    \centering\footnotesize\textbf{(g) \Prefilled}\\completed scaffold
  \end{minipage}\hfill
  \begin{minipage}[t]{0.22\textwidth}
    \vspace{0pt}
    \includegraphics[width=\linewidth,trim=0 270 0 0,clip]{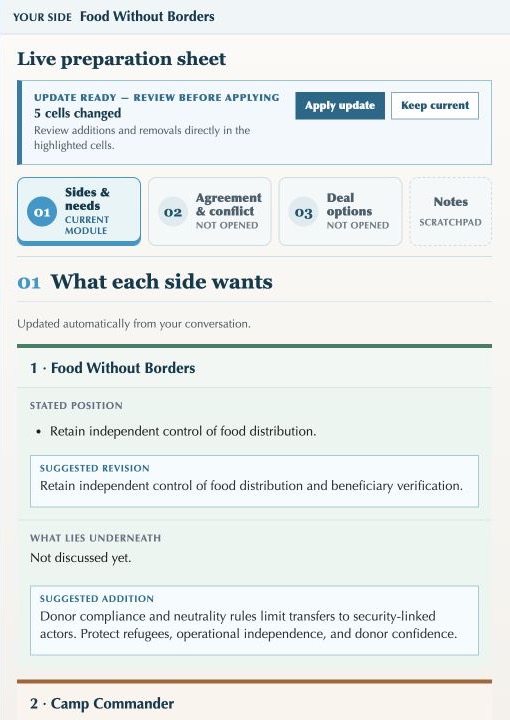}
    \centering\footnotesize\textbf{(h) \Coevolving}\\proposed update
  \end{minipage}
  \caption{The four preparation interfaces and their condition-specific components. Panels (a)--(d) show the complete interfaces during the reading phase; panels (e)--(h) isolate the components that distinguish them. All conditions shared the same document list and reader. \Reader{} provided private notes, \Chatbot{} added a document-aware assistant, \Prefilled{} presented a completed analysis organized by the professional scaffold, and \Coevolving{} began with the same scaffold empty and later presented conversation-derived changes for participants to apply or reject.}
  \Description{Eight interface panels. Four large overview panels show the complete Reader, Chatbot, AI-Prefilled, and Co-Evolving interfaces. Four detail panels show the Reader notes field, the Chatbot request field, the AI-Prefilled completed preparation scaffold, and the Co-Evolving preparation scaffold with highlighted suggested changes and controls to apply the update or keep the current version.}
  \label{fig:interfaces}
\end{figure*}

We conducted a four-condition between-subjects study comparing two interfaces, \Prefilled, and \Coevolving, against two baselines, \Reader and \Chatbot, on a realistic frontline humanitarian negotiation preparation task (Figure~\ref{fig:interfaces}). \Reader served as the no-AI baseline and presented the case documents alongside a note-taking area. \Chatbot served as the AI baseline with a conversational assistant below the reader. \Prefilled presented an AI-generated, scaffolded analysis of the case before participants began their preparation. The scaffold follows three standard preparation frameworks from the CCHN Field Manual that frontline negotiators use to prepare cases: the Iceberg, the Island of Agreements, and Paths to Agreement~\cite{cchn}. We adapted these frameworks for the purpose of the study such that they are succinct enough for this short study. \Coevolving combined the \Chatbot with a scaffold that was progressively updated based on the participant's conversation with the AI.

\subsection{Participants}
The study was approved by the Institutional Review Board at our institution. 

We recruited adult participants through Prolific; recruitment did not target professional or expert negotiators. Participants were required to complete the study on a desktop or laptop computer, and the interface enforced a minimum viewport size of 1024 by 600 pixels. All participants provided informed consent before beginning the study and were compensated \Pay, corresponding to an estimated hourly rate of \Payrate. Condition was randomized, concealed until the interface loaded, and stable for any participant who re-entered. Table~\ref{tab:participants} reports the realized $n$ per condition.

We excluded participants who left the study mid-session and rejoined, returning workers through Prolific, and participants who reported at intake that they had taken the study before. The study contained no separate attention-check item. To limit the use of outside AI tools, the interface blocked \texttt{copy}, \texttt{cut}, and \texttt{paste} events, drag-and-drop, and the context menu on every task screen; these were re-enabled only on the written-response screens, where a paste into an answer field was logged with its target field and character count. Keyboard shortcuts associated with screen capture (\texttt{PrintScreen} variants, \texttt{Cmd+Shift+3/4/5}, and snipping-tool chords) were intercepted during the timed preparation phase and replaced with an on-screen notice; after three attempts the session ended, the participant was routed to a policy-exit screen, and their data were excluded from analysis.

\subsection{Procedure}

\subsubsection{Demographics Survey} Before beginning the task, participants reported their age, gender, computer-use frequency, generative-AI-tool use, English proficiency, educational attainment, country of childhood, current country of residence, prior negotiation experience, and whether they had previously taken the study. The complete intake instrument appears in Appendix~\ref{app:surveys}.

\subsubsection{Interface Tutorial}
Before the timed preparation phase, participants completed a non-skippable tutorial using a neutral practice case unrelated to the experimental scenario. The tutorial reproduced the interface assigned to the participant but used predefined content and made no live AI calls. The \Tprep-minute preparation timer began only after the tutorial was completed. Appendix~\ref{app:tutorial} reproduces the practice documents and participant-facing tutorial text.

\subsubsection{Preparation Phase}

The first major task was the preparation task. In this task, participants were presented with \Nfiles{} negotiation case files (Appendix~\ref{app:case-files}). One frontline negotiator with more than ten years of experience and two members of the research team jointly developed the case files used in the study. The scenario was adapted from a realistic case in the \textit{CCHN Field Manual}, a practitioner handbook for frontline humanitarian negotiation~\cite{cchn}. The \Tprep-minute window was intentionally short to approximate frontline settings in which negotiators must prepare with incomplete information and adapt quickly to changing operational conditions, including sudden security incidents~\cite{cchn,ma2025chatgpt}.

Participants represented FWB, a fictitious international humanitarian organization seeking to deliver urgently needed food assistance to a refugee camp in Country Alpha. They prepared to negotiate with the Camp Commander, a local militia member who controlled access through the camp administration and its guard structures. Both parties sought rapid food distribution but disagreed over its terms. FWB required an independent nutritional assessment, control over beneficiary selection and distribution, and independently vetted civilian workers; it prohibited resource transfers to militia-linked actors. The Commander demanded immediate distribution using existing registration data, guard-led operations, restricted and escorted access for FWB staff, and food rations for militia members and their families as compensation. Participants therefore had to develop an arrangement that addressed access, security, and distribution while preserving FWB's neutrality and operational independence.

Participants encountered these positions across 15 case documents totaling 6,977 words, rather than as a consolidated summary. The corpus included role cards, operational and logistics communications, negotiation minutes, a donor compliance Q\&A, and messages from refugee and host-community representatives. It combined direct records with conflicting or incomplete claims, requiring participants to distinguish documented constraints from assertions and integrate information across sources. All participants received the same corpus, listed in Appendix~\ref{app:case-files}.

\subsubsection{Mid-survey}

Immediately after the \Tprep-minute preparation phase, participants completed three brief self-report measures before beginning the assessment task. They rated the mental effort required by the preparation task, their confidence that they had prepared effectively, and their confidence in entering the negotiation with a clear strategy. All three items used seven-point scales ranging from \textit{strongly disagree} to \textit{strongly agree}. Appendix~\ref{app:surveys} gives the item wording.

\subsubsection{Assessment Phase}
\label{sec:assessment}
Participants then completed ten free-text questions assessing their understanding of the case and their preparation for the negotiation. Participants were informed that concise bullet points and incomplete sentences were acceptable. Appendix~\ref{app:questions} reproduces the instructions and questions. 

The assessment phase lasted \Tassess{} minutes and was divided into four sections:
\begin{enumerate}
    \item \textit{Participant's own side} (Questions 1--3): negotiated issues, FWB's objectives, and FWB's redlines (5 minutes);
    \item \textit{Opposing side} (Questions 4--5): the Camp Commander's objectives and redlines (3.5 minutes);
    \item \textit{Overlap and conflict} (Questions 6--7): common ground and major conflicts (3.5 minutes); and
    \item \textit{Scenario planning} (Questions 8--10): a realistic package, risks to the package, and clarification questions (5 minutes).
\end{enumerate}

Participants could advance before a section timer expired, in which case unused time was added to the following section. Participants could not return to an earlier section after advancing. If a section timed out, the study automatically advanced to the next section.

The full case files were no longer available during writing. Instead, participants could consult the materials retained from their assigned preparation interface. \Reader participants retained their notes; \Chatbot participants retained their notes and conversation history; \Prefilled participants retained the scaffolded preparation artifact and guide transcript; and \Coevolving participants retained the evolving artifact and their complete conversation history. These materials were displayed beside the questions as reference material.

\subsubsection{Post-survey}
After completing the assessment phase, participants completed an untimed post-task survey. The survey first asked whether participants had prior familiarity with CCHN or other negotiation frameworks. It then measured participants' perceived decision agency, psychological ownership, case understanding, perceived usefulness of their preparation, mental effort, and frustration. Appendix~\ref{app:surveys} reproduces the complete instrument.

Decision agency was measured with three items assessing whether participants felt that they controlled the preparation process, could choose which information or suggestions to use, revise, or ignore, and felt constrained by the system. The system-constraint item was reverse-coded before aggregation. Psychological ownership was measured with three items assessing whether the preparation felt like the participant's own work, whether the participant felt responsible for its content, and whether it reflected the participant's own reasoning.

Participants also rated whether the preparation process helped them understand the case and whether the resulting preparation would be useful when entering the negotiation. Finally, they reported how hard they had worked mentally to answer the written questions and how frustrating it was to use the available information. All subjective measures used seven-point response scales. Agreement items ranged from \textit{strongly disagree} to \textit{strongly agree}, whereas mental-effort and frustration items ranged from \textit{very low} to \textit{very high}. Appendices~\ref{app:questions} and~\ref{app:surveys} reproduce the written questions and study instruments.

\subsection{Preparation Interfaces}
In both scaffolded conditions, a separate evidence-highlighting pass marked case-file passages relevant to the active scaffold section. Participants could save a highlighted passage to their notes with one click and inspect it in context in the case reader.

Participant-facing chat in \Chatbot, \Prefilled and \Coevolving used the same model (\texttt{gpt-5.6-luna}, reasoning effort \emph{none}). The prefilled sheet in \Prefilled and the conversation-to-artifact extractor in \Coevolving used \texttt{gpt-5.6-terra}; evidence highlighting used \texttt{gpt-5.4-mini}. Because the prefilled cards and the evidence scans depend only on the prompt, model, role, and file content, they were content-cached, so every \Prefilled participant received an identical prepared sheet. Outputs that depend on participant input were generated live and therefore differ across participants by design. Appendix~\ref{app:system} provides additional implementation details for the four interfaces. 

\subsubsection{Reader}

The \Reader condition (Figure~\ref{fig:interfaces}a and e) served as the no-AI baseline. Participants navigated the case files through a numbered document list and read each file in a central reading pane. A notes panel allowed participants to record information in their own words. During the assessment phase, participants could refer to the notes they had created.

\subsubsection{Chatbot}

The \Chatbot condition (Figure~\ref{fig:interfaces}b and f) added a conversational assistant to the \Reader interface. The assistant received the complete case corpus as context. Participants could ask questions about the case in natural language or use an in-interface control to attach a selected passage to the conversation. Their notes and complete chat history remained available during the assessment phase.

\subsubsection{AI-Prefilled}

Open-ended chat preserves flexibility, but it presents analysis as a linear sequence of exchanges and leaves users to determine what a complete preparation should contain. We therefore introduced a persistent preparation artifact that made the professional components of the task visible.

Thus, the \Prefilled condition contains an AI-generated preparation artifact organized by the scaffold in addition to the \Chatbot (Figure~\ref{fig:interfaces}c and g). The \Prefilled{} condition tested whether showing participants a completed analysis organized by a professional scaffold improved preparation even when they did not need to formulate prompts or construct the initial synthesis.

Before participants began interacting with the materials, the AI populated three modules, each corresponding to one CCHN preparation framework: the Iceberg (stated positions and the concerns beneath them), the Island of Agreements (shared ground and core tensions), and Paths to Agreement (a workable package, with boundaries and redlines recorded alongside it)~\cite{cchn}. We adapted the negotiation frameworks to make them succinct for a short \Tprep-minute preparation phase. In the AI-prepared material, each claim included citations that participants could use to inspect the supporting case file, which could be saved as notes. Participants could navigate and verify the prepared analysis but were not required to formulate prompts or construct the initial synthesis themselves. They could also submit a free-text instruction asking the AI to revise one module. The preparation artifact and chatbot transcript remained available during the assessment phase.

\subsubsection{Co-Evolving}

The \Coevolving condition combined the chatbot with the same three-module preparation scaffold used in the \Prefilled condition. Unlike the \Prefilled interface, however, the artifact began empty. Participants first optionally explored the case through natural-language conversation with the \Chatbot. As the conversation developed, the system organized the positions, interests, areas of agreement and conflict, proposed packages, implementation constraints, and risks raised in that conversation into the corresponding artifact cells.

The first conversation automatically initialized relevant portions of the artifact. Subsequent changes appeared as revisions to affected cells, which participants could apply or reject while retaining the current version (Figure~\ref{fig:interfaces}h). The interface also highlighted one area that remained underdeveloped and provided a short prompt to the user indicating what was known and what had not yet been discussed. The highlighting followed the Naivasha Grid, a negotiation preparation checklist in the CCHN Field Manual~\cite{cchn}. A language model judged which part of the grid remained underdeveloped. Appendix~\ref{app:system} describes the update and highlighting logic.

\subsection{Scoring Negotiation Preparation}

We scored each participant's written preparation against a rubric in appendix~\ref{app:rubric}. Two frontline negotiators derived and froze the rubric definitions from the case record. We then used an LLM-as-a-judge to apply them consistently across responses~\cite{zheng2023judging}.

\subsubsection{Deriving the Criteria}
Two frontline negotiators constructed the rubric directly from the case files. For Questions 1--7 and 9, they identified atomic concepts that a well-prepared negotiator could draw from the record. Question 8 required participants to construct an agreement package, so the negotiators instead defined seven package issues and a closed option set for each issue.

Each question was scored against a question-specific set of coverage items, with each item representing one distinct element of a complete answer. For example, the answer to FWB-redlines question included separate items for control over beneficiary selection and distribution, prohibition on transfers to militia-linked actors, and use of independently vetted civilian workers. Appendix~\ref{app:rubric} summarizes the scoring structure and its validation against two expert raters.

\subsubsection{Coverage Measures}
\label{sec:coverage-measures}
For Questions 1--7 and 9, coverage was the proportion of question-specific items supported by the answer. For Question 8, which asked participants to propose a package to the counterparty, the judge coded how each of seven package issues was addressed. An issue counted as covered if the package mentioned an option for that issue and as uncovered if it did not. Overall preparation coverage was the equally weighted mean of the nine question-level coverage scores.

During rubric development, the negotiators concluded that clarification questions (Question 10) did not admit a bounded scoring because what required clarification depended on each participant's proposed package and remaining uncertainties. Question 10 was therefore retained as an open-ended response and was not included in the analysis.

This procedure gives each scored question equal weight, regardless of the number of criteria or package issues it contains. The two negotiators independently coded a condition-balanced reference set of 100 participants before resolving disagreements. We then validated candidate LLM judges against the adjudicated reference set using a participant-disjoint split. The resulting grader used \texttt{gpt-5.6-sol} with no reasoning effort. Appendix Table~\ref{tab:grading-validation} reports expert agreement and held-out grader performance.

\subsubsection{Proposed Package Measurement}

The proposed-package question required additional evaluation because mentioning an issue did not establish which risks the proposed arrangement introduced. We therefore evaluated Question 8 against an option table developed by two frontline negotiators from the case record. The table contained 29 possible proposals across the seven issues of the case: camp security, humanitarian-operations governance, compensation, access and movement, assessment and timing, implementer selection, and host-community pressure.

For each option, the negotiators identified the risks associated with that arrangement. After the option table was frozen, a separate rubric-grounded LLM judge mapped the Question 8 answer directly to one option per issue without seeing its risk labels. \textit{Not addressed} marked an issue as uncovered. Appendix Table~\ref{tab:grading-validation} reports held-out validation of issue coverage.

Question 9, proposed package's risks, contains two measures. Risk coverage was the proportion of case-grounded risk criteria identified. Risk relevance was the share of a participant's identified risks that applied to the options resolved in that participant's proposed package. Thus, risk coverage measures how broadly participants anticipated failure modes, whereas risk relevance measures whether those risks matched the package they had actually proposed.

We tested three nested contrasts: all AI-supported conditions versus \Reader{}, the two scaffolded conditions versus \Chatbot{}, and \Coevolving{} versus \Prefilled{}. Overall coverage used Welch tests. Package-quality $p$-values used Benjamini--Hochberg correction across risk coverage and risk relevance within each contrast.

\subsubsection{Prompt Purposes}

We characterized how participants used conversational AI by classifying all 2{,}992 participant turns sent to an assistant in the three AI conditions. Each turn received exactly one \textit{request-form} code and one or more \textit{purpose} codes. Request form distinguished participant-authored messages, selected passages accompanied by an authored request, and selected passages sent without an authored request. For example, highlighting a passage and asking AI to explain it was coded as a passage plus request; sending only the highlighted passage was coded as passage only.

The 12 purposes were reading support, fact retrieval, analysis of FWB's position, analysis of the Camp Commander's position, cross-party synthesis, negotiation strategy or package development, deliverable drafting, roleplay, verification or challenge, interaction control, task or interface questions, and other purposes (Appendix~\ref{app:code-frame}). 

We established an adjudicated human reference set of 400 condition-balanced messages, each drawn from a different participant. Two co-authors independently coded the turns while blind to study condition, then resolved disagreements to establish the reference labels. Agreement and classifier validation appear in Appendix Tables~\ref{tab:prompt-classifier-validation}--\ref{tab:prompt-program-selection}.

Our primary analyses measured whether each participant used AI for each purpose at least once. We distinguished four analytic purposes: \emph{own-side analysis}, interpreting FWB's interests, constraints, or risks (e.g., ``How do FWB's redlines limit the concessions it can offer?''); \emph{counterpart analysis}, interpreting the Camp Commander's interests, constraints, or likely reactions (e.g., ``Why might the Commander resist using independently vetted civilian workers?''); \emph{cross-party synthesis}, comparing or integrating the two parties' positions (e.g., ``Where do FWB's and the Commander's interests align, and where do they conflict?''); and \emph{strategy or package development}, recommending negotiation actions or constructing proposed terms (e.g., ``What package could FWB propose to address the Commander's security concerns without crossing its own redlines?''). The categories captured interpretation, comparison, or recommended action rather than simply retrieving or restating case facts.

Analytic breadth was the number of these four purposes observed across a participant's turns, ranging from zero to four. To compare breadth at a common interaction volume, we also analyzed participants with at least three assistant interactions. We repeatedly sampled three interactions from each participant and averaged the number of distinct analytic purposes represented in those samples.

We first tested whether participants used the assistant at all. We used Pearson's chi-squared test for the overall condition difference and Holm-corrected Fisher’s exact tests for pairwise comparisons. Among assistant users, we compared the incidence of each purpose using participant-level differences in proportions with HC3 confidence intervals. The scaffolded contrast combined \Prefilled{} and \Coevolving{}, weighted by their observed numbers of assistant users. We applied Benjamini--Hochberg correction across the 12 purposes within each contrast and Holm correction across the three measures of analytic breadth. Finally, we examined whether preparation quality was associated with how participants used the assistant. For each purpose, we regressed preparation quality on purpose incidence, condition, and log participant-turn count among assistant users. Thus, the estimated association compares participants in the same condition with similar interaction volumes. We report HC3 confidence intervals and Benjamini--Hochberg-adjusted results across the 12 purpose models. 

%% file: sections/05-results.tex
\section{Results}
\label{sec:results}

\subsection{Sample and attrition}

Of \Nrecruited{} participants who entered the study, \Nfinal{} completed it, provided a final survey, and met the inclusion criteria. The analyzed sample comprised \Nreader{} participants in \Reader{}, \Nchatbot{} in \Chatbot{}, \Nprefilled{} in \Prefilled{}, and \Ncoevolving{} in \Coevolving{} (Table~\ref{tab:participants}).

Completion differed by condition ($p=.007$), but the observed characteristics of analyzed participants did not differ detectably across conditions (all $p\geq.17$; Table~\ref{tab:participants}). The results below are complete-case estimates. Sessions remained in the analysis when the written-response timer expired.

\begin{table*}
\centering
\caption{Sample disposition and observed participant characteristics by condition. Completion is calculated among assigned participants; all other rows describe the analyzed sample.}
\label{tab:participants}
\small
\begin{tabular*}{\textwidth}{@{\extracolsep{\fill}}lrrrr@{}}
\toprule
& \Reader{} & \Chatbot{} & \Prefilled{} & \Coevolving{} \\
\midrule
Analyzed, $n$ & 181 & 260 & 206 & 153 \\
Completed assigned study & 48.3\% & 54.6\% & 45.1\% & 44.9\% \\
Written-response timer expired & 35.4\% & 39.6\% & 43.7\% & 37.3\% \\
\midrule
Age, $M$ (SD) & 38.2 (13.1) & 38.4 (12.3) & 37.2 (11.7) & 37.9 (11.6) \\
Woman & 59.7\% & 55.0\% & 52.9\% & 52.3\% \\
Bachelor's or higher & 62.9\% & 64.2\% & 66.5\% & 60.8\% \\
Native English & 93.9\% & 93.8\% & 92.2\% & 92.8\% \\
No negotiation experience & 52.5\% & 48.8\% & 42.7\% & 44.4\% \\
Uses generative AI daily & 51.4\% & 54.6\% & 52.9\% & 56.9\% \\
\bottomrule
\end{tabular*}
\end{table*}

\subsection{AI Improved Preparation Coverage}
\label{sec:results-quality}
\label{sec:accuracy}
\label{sec:results-effort}
\label{sec:results-ownership}

\begin{figure*}
\centering
\includegraphics[width=\textwidth]{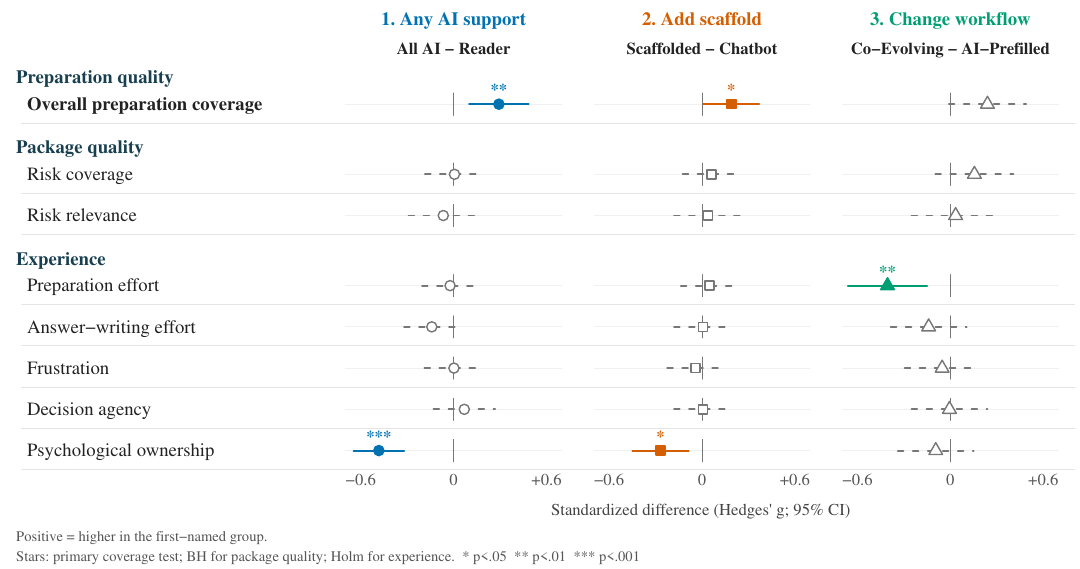}
\caption{\textbf{AI improved preparation coverage; the visible scaffold added a further coverage gain.} \Coevolving{} required less reported preparation effort than \Prefilled{}, while AI-supported conditions reported lower psychological ownership than \Reader{}. Each panel reports Hedges' $g$ with 95\% confidence intervals; exact condition summaries and native-unit estimates appear in Appendix Table~\ref{tab:overall-outcomes}. Positive estimates indicate a higher value in the first-named group. Overall coverage uses Welch tests; package-quality tests use Benjamini--Hochberg correction within each contrast, and experience measures use Holm correction across the three contrasts within each outcome. $^{*}p<.05$, $^{**}p<.01$, $^{***}p<.001$.}
\Description{Three aligned forest plots compare all AI conditions with \Reader{}, both scaffolded conditions with \Chatbot{}, and \Coevolving{} with \Prefilled{}. AI improves overall preparation coverage, and scaffolding adds a further coverage gain. \Coevolving{} has lower preparation effort than \Prefilled{}. AI is associated with lower psychological ownership, while risk outcomes and decision agency change little.}
\label{fig:nested-outcome-comparisons}
\end{figure*}

AI support increased average preparation coverage by 3.65 points relative to \Reader{} (95\% CI [1.25, 6.05], $p=.003$, $g=0.25$; Figure~\ref{fig:nested-outcome-comparisons}), supporting H1. Making the professional scaffold visible added a further 2.36 points over \Chatbot{} (95\% CI [0.11, 4.61], $p=.040$, $g=0.16$), supporting H2. The estimated advantage of \Coevolving{} over \Prefilled{} was 3.19 points, but was less precise (95\% CI [$-0.14$, 6.52], $p=.060$, $g=0.21$; Appendix Table~\ref{tab:overall-outcomes}).

Risk coverage and the match between identified risks and the proposed package did not differ detectably in any nested contrast (all $p_{\mathrm{BH}}\geq.45$; Figure~\ref{fig:nested-outcome-comparisons}).

Among the subjective measures, preparation effort distinguished the two scaffolded workflows. \Coevolving{} participants reported less effort than \Prefilled{} participants (6.17 versus 6.55, $g=-0.35$, $p_{\mathrm{Holm}}=.005$), while the other contrasts for preparation effort, answer-writing effort, and frustration were not detectable (all $p_{\mathrm{Holm}}\geq.41$; Figure~\ref{fig:nested-outcome-comparisons}; exact summaries in Appendix Table~\ref{tab:overall-outcomes}). Incrementally developing the scaffold thus required less reported preparation effort than reading a completed AI analysis.

AI conditions' psychological ownership reduced compared to \Reader{} ($g=-0.42$, $p_{\mathrm{Holm}}<.001$) and lower across the scaffolded conditions than in \Chatbot{} ($g=-0.23$, $p_{\mathrm{Holm}}=.015$), with no detectable difference between \Coevolving{} and \Prefilled{} ($g=-0.08$, $p_{\mathrm{Holm}}=.49$). Decision-agency means remained between 5.04 and 5.12 in every condition, and none of the nested contrasts was detectable (all $p_{\mathrm{Holm}}=1.00$; Figure~\ref{fig:nested-outcome-comparisons}).

\subsection{Preparation Gains Spanned Factual Coverage and Package Proposals}
\label{sec:results-selectivity}

The gains spanned factual coverage and package proposals. Relative to \Reader{}, the AI-supported conditions improved coverage of the Commander's objectives (+5.44 points), common ground (+5.78), main conflicts (+6.06), and package issues (+7.94; all $q\leq.0014$). Among participants with AI access, the scaffolded conditions exceeded \Chatbot{} on the Commander's redlines (+5.75 points, $q=.0066$) and main conflicts (+7.91, $q<.001$; Appendix Figure~\ref{fig:question-quality-comparisons}). Package-risk coverage did not change detectably ($q=.95$), and none of the nine question-level \Coevolving{}--\Prefilled{} contrasts survived correction.

\subsection{Scaffolding Shifted AI Use toward Negotiation Analysis}
\label{sec:results-prompts}

Assistant usage depended on the workflow, $\chi^2(2)=44.98$, $p<.001$: 81.9\% of \Chatbot{} participants, 62.6\% of \Prefilled{} participants, and 90.8\% of \Coevolving{} participants interacted with it at least once. Assistant usage was lower in \Prefilled{} than in \Chatbot{} ($p_{\mathrm{Holm}}<.001$), and higher in \Coevolving{} than in either \Chatbot{} ($p_{\mathrm{Holm}}=.014$) or \Prefilled{} ($p_{\mathrm{Holm}}<.001$). The purpose analyses below describe these 481 assistant users (exact summaries in Appendix Tables~\ref{tab:prompt-intents} and~\ref{tab:assistant-use-comparisons}).

\begin{figure*}
\centering
\includegraphics[width=\textwidth]{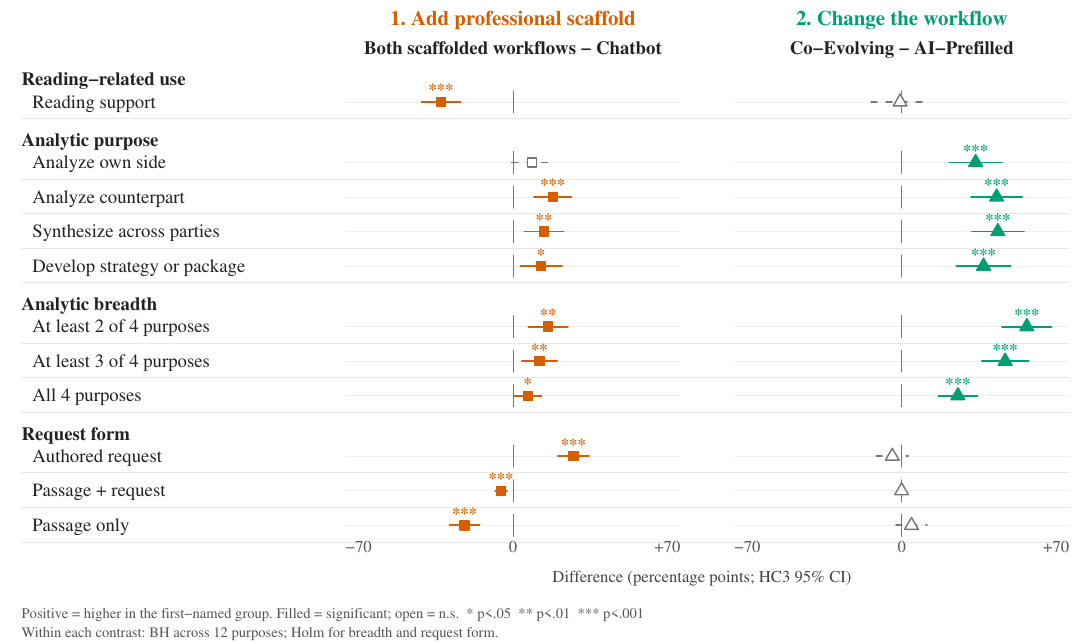}
\caption{\textbf{Scaffolding shifted AI use away from reading support, while \Coevolving{} elicited a broader analytic repertoire than \Prefilled{}.} Points are differences in percentage points with HC3 95\% confidence intervals. Purpose tests use Benjamini--Hochberg correction across all 12 purposes; analytic-breadth and request-form tests use Holm correction within each contrast. Request-form rows report the mean within-participant share of turns. Filled colored markers, solid intervals, and stars indicate adjusted $p<.05$; open gray markers and dashed intervals are nonsignificant.}
\Description{Two aligned forest plots compare both scaffolded workflows with \Chatbot{} and \Coevolving{} with \Prefilled{}. Scaffolding reduces reading support and passage-only turns while increasing several analytic purposes. \Coevolving{} exceeds \Prefilled{} on all four analytic purposes and all three breadth thresholds, while request form is similar.}
\label{fig:prompt-use-patterns}
\end{figure*}

The visible scaffold changed what participants asked AI to do. Relative to \Chatbot{}, the scaffolded workflows reduced reading-support use by 30.1 percentage points ($q<.001$), while increasing counterpart analysis by 16.6 points ($q<.001$), cross-party synthesis by 13.0 points ($q=.006$), and strategy or package development by 11.8 points ($q=.020$; Figure~\ref{fig:prompt-use-patterns}).

The two scaffolded workflows differed in analytic breadth. Relative to \Prefilled{}, \Coevolving{} increased the incidence of each analytic purpose by 30.9 to 40.2 percentage points (all $q<.001$), the share of participants covering at least two purposes by 52.3 points, and the share covering all four by 23.5 points (both adjusted $p<.001$; Figure~\ref{fig:prompt-use-patterns}). The difference remained when interaction volume was held constant: among participants with at least three turns, the expected number of analytic purposes in a three-turn sample was 1.68 in \Coevolving{}, versus 0.91 in \Chatbot{} and 0.88 in \Prefilled{} (both \Coevolving{} comparisons $p_{\mathrm{Holm}}<.001$; Appendix Table~\ref{tab:prompt-intents}).

Scaffolding shifted participants from submitting passages to authoring requests, but explicit verification remained uncommon. In the scaffolded workflows, participant-authored requests comprised 25.2 percentage points more of each participant's turns than in \Chatbot{}, while passage-only turns comprised 20.2 points less (both adjusted $p<.001$; Figure~\ref{fig:prompt-use-patterns}); \Coevolving{} and \Prefilled{} did not differ detectably on these measures. Yet just 72 of 2{,}992 turns (2.4\%) asked AI to check, defend, correct, or support a claim, representing 52 of 481 assistant users (10.8\%; Appendix Figure~\ref{fig:prompt-purpose-inventory}).

\subsection{Analytic AI Use Tracked Stronger Preparation}
\label{sec:results-prompt-quality}

\begin{figure}[t]
\centering
\includegraphics[width=0.62\textwidth]{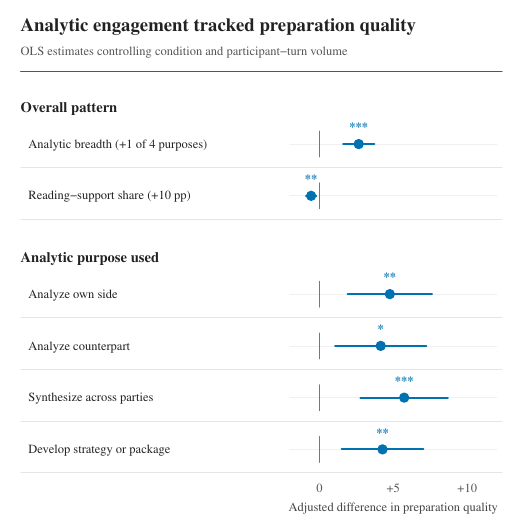}
\caption{\textbf{Analytic engagement was associated with stronger preparation, whereas a larger reading-support share was associated with lower coverage.} Exploratory OLS estimates use the 481 participants who interacted with the assistant, the updated equal-question Questions 1--9 outcome, and controls for condition and log participant-turn count; intervals use HC3 standard errors. Individual-purpose stars use Benjamini--Hochberg correction across 12 purposes.}
\Description{A coefficient plot shows a positive association between analytic breadth and preparation coverage and a negative association for reading-support share. Each of the four analytic purposes has a positive adjusted association with coverage.}
\label{fig:prompt-quality-associations}
\end{figure}

Among assistant users, a broader analytic repertoire was associated with stronger preparation. In an exploratory OLS model controlling for condition and log participant-turn count, each additional analytic purpose corresponded to 2.66 more coverage points (95\% CI [1.60, 3.72], $p<.001$), whereas a ten-percentage-point increase in reading-support share corresponded to 0.56 fewer points (95\% CI [$-0.91$, $-0.21$], $p=.002$; Figure~\ref{fig:prompt-quality-associations}). Own-side analysis (+4.77 points, $q=.004$), counterpart analysis (+4.15, $q=.021$), cross-party synthesis (+5.74, $q<.001$), and strategy or package development (+4.27, $q=.008$) each had positive associations with coverage. In a joint model, their average coefficient was +2.65 points (95\% CI [1.59, 3.72], $p<.001$), with no detectable differences among the four coefficients ($F(3,473)=0.32$, $p=.814$).

\subsection{Prior Experience Was Associated with Higher Performance with No AI-Support, but Not AI-Supported, Preparation}
\label{sec:results-experience}

\begin{figure*}
\centering
\includegraphics[width=\textwidth]{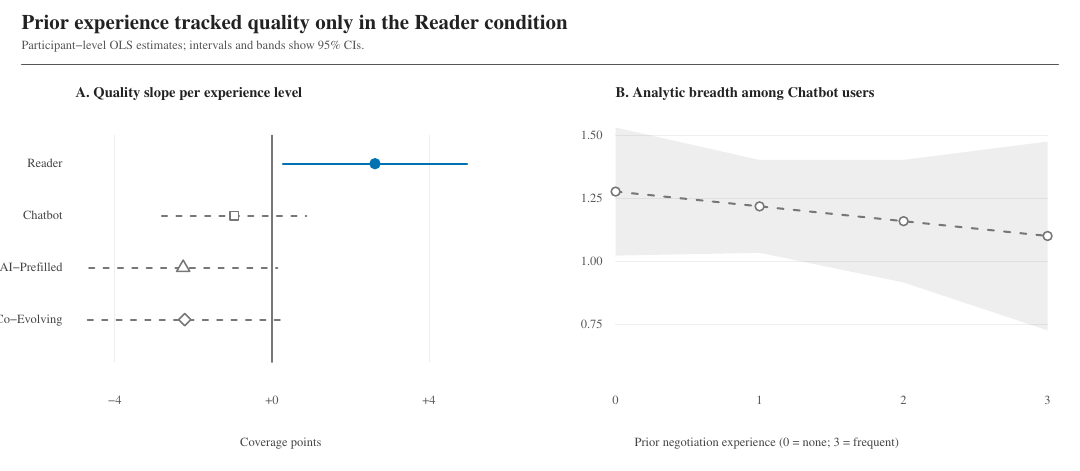}
\caption{\textbf{Prior experience tracked preparation quality in \Reader{}, but not detectably in the AI conditions; it was also not associated with analytic breadth in \Chatbot{}.} Panel A shows the condition-specific change in preparation-quality coverage for each step on the four-level experience scale. Panel B shows estimated four-purpose analytic breadth among \Chatbot{} participants who interacted with the assistant. Exploratory OLS intervals use HC3 standard errors.}
\Description{Panel A shows a positive association between prior experience and preparation quality in \Reader{}, while the three AI-condition slopes cross zero. Panel B shows a nearly flat association between experience and analytic breadth among \Chatbot{} users.}
\label{fig:experience-moderation}
\end{figure*}

Prior experience was associated with preparation quality in \Reader{}, where each step on the four-level scale corresponded to 2.62 additional coverage points (95\% CI [0.27, 4.97], $p=.029$). The corresponding slopes were $-0.96$ in \Chatbot{}, $-2.26$ in \Prefilled{}, and $-2.22$ in \Coevolving{} (all $p\geq.065$). The \Reader{} slope exceeded the participant-weighted mean slope across the AI conditions by 4.33 points (95\% CI [1.66, 7.00], $p=.002$; Figure~\ref{fig:experience-moderation}).

Nor did prior experience predict a broader analytic repertoire in open chat. Among \Chatbot{} participants who used the assistant, each step in experience was associated with $-0.06$ analytic purposes (95\% CI [$-0.23$, 0.11], $p=.49$; Figure~\ref{fig:experience-moderation}).

\FloatBarrier

%% file: sections/06-discussion.tex
\section{Discussion}
\label{sec:discussion}
In a four-condition experiment, we showed that AI-supported interfaces improved negotiation preparation coverage over reading the case without AI assistance (Figure~\ref{fig:nested-outcome-comparisons}). Pooling the two scaffold interfaces, \Prefilled{} and \Coevolving{}, improved coverage further than open-ended chat. We also found an echo of a result reported in other domains: prior negotiation experience predicted preparation quality when participants worked without AI, and it stopped predicting it under all three AI interfaces (Figure~\ref{fig:experience-moderation}). Although no professional negotiators were specifically recruited for this study, a participant who had never negotiated identified about as much of the case as a participant who had more experience. Generative AI has been found to raise the immediate performance of less experienced workers the most~\cite{brynjolfsson2025generative,dellacqua2026jagged}.

\subsection{Open-Ended Chatbots Do Not Remove Designers' Responsibility to Scaffold the Work}
Open-ended chatbots are often treated as an alternative to task-specific interface design. Instead of anticipating what users need, designers can expose a general conversational interface and allow users to request the appropriate assistance~\cite{tankelevitch2024metacognitive,subramonyam2024gulf}. Our findings challenge this premise. Although chat makes many forms of assistance technically available, it leaves users responsible for recognizing what matters, decomposing the task, and formulating productive requests. For example, participants using \Chatbot{} predominantly asked the chatbot to summarize or explain the materials, leaving much of its analytic potential unused. Prior negotiation experience did not produce a reliably different prompting repertoire either. More experienced participants and novices initiated chat at similar rates and asked similar types of questions. This complements prior work showing that non-AI experts tend to explore prompts opportunistically, overgeneralize from isolated outputs, and struggle to evaluate whether a prompting strategy is robust~\cite{zamfirescu2023johnny}. Experience may help people reason about a negotiation, as it did in \Reader{}, but it does not necessarily teach them how to translate that reasoning into productive requests to  an AI.

Together, these results suggested that open-ended chat shifts substantial interaction-design work onto users regardless of user domain expertise. Natural-language flexibility therefore does not necessarily provide task guidance. It may simply leave users responsible for supplying the workflow that the interface misses.

Recent work on generative and malleable interfaces might appear to reduce this burden by allowing the interface itself to adapt to a user’s requests or activity~\cite{leviathan2026generativeui,lam2026justintime,vaithilingam2024dynavis,cao2025malleable}. Such systems can generate task-specific controls from a prompt, infer a user's current objective and surface relevant tools, or transform natural-language requests into persistent, manipulable interface elements. For example, DynaVis turns a request to modify a visualization into an editable visual control rather than a one-time textual instruction \cite{vaithilingam2024dynavis}. However, generative and malleable interfaces do not eliminate the prior question of which objectives the interface should support. If interface adaptiveness is driven primarily by what users request or by what a system infers from their immediate activity, consequential dimensions of the professional task may remain unsupported.

Domain knowledge can provide a basis for deciding which objectives an adaptive interface should surface rather than leaving those objectives to users or the system to discover. Compared to \Chatbot{}, \Coevolving{} surfaced preparation objectives derived from negotiation frameworks, elicited a broader analytic repertoire, and shifted requests away from document processing toward analytic questions associated with stronger preparation (Figures~\ref{fig:prompt-use-patterns} and~\ref{fig:prompt-quality-associations}). At the same time, a scaffold directs attention toward the conception of the task that it represents. The scaffold improved several factual and package dimensions, but it did not produce a detectable gain in package-risk coverage (Appendix Figure~\ref{fig:question-quality-comparisons}). Participants also rarely asked the AI to justify or verify its suggestions. Scaffolding can make important analytic work visible while leaving other consequential work implicit.

The design challenge, then, is to provide a useful scaffold without treating it as a complete representation of the task. One approach is to make the scaffold itself inspectable and revisable. For example, DISCERN lets line managers construct and revise a value tree representing the objectives and criteria relevant to a workplace decision, while synchronizing those changes with the information they record and compare~\cite{khadpe2024discern}. Such approaches allow a domain-informed scaffold to guide the work without making it fixed or exhaustive. Designers should therefore treat scaffolds as hypotheses about what the work requires and provide ways to revise what those scaffolds make visible as the work unfolds.

\subsection{Designing AI for Outcomes Beyond Performance}
Much of the emerging literature on generative AI at work evaluates its value through improvements in output quality, completion time, and productivity
. Across writing, customer support, consulting, and other knowledge-work tasks, AI assistance has enabled workers to produce better outputs, complete tasks faster, or both~\cite{noy2023productivity,brynjolfsson2025generative,dellacqua2026jagged}. These measures capture important benefits of AI-supported work, but they do not reveal what kind of work workers perform to achieve those outcomes or how workers experience that work. AI-supported workflows may differ not only in performance, but also in the cognitive effort they require, the agency workers retain over decisions, and the ownership workers feel over what they produce. Our results show why these outcomes need to be considered separately.

Improved performance did not translate into perception of reduced effort. AI improved preparation quality, but none of the AI-supported conditions reduced subjective effort relative to \Reader{}. \Prefilled{} produced the highest preparation-effort ratings and was experienced as more demanding than \Coevolving{}. \Chatbot{} and \Coevolving{} were likewise no less demanding than \Reader{}, and effort during the written-answer phase did not differ across conditions. In this task, AI redistributed cognitive work rather than simply removing it. Receiving a comprehensive AI-generated analysis required \Prefilled{} participants to read additional AI material other than the case file at the beginning. \Chatbot{} and \Coevolving{}, by contrast, presented information through a sequence of exchanges. Their lower effort relative to \Prefilled{}, as we hypothesize, is consistent with participants being able to control the pace of analysis and process it incrementally. An AI system should therefore not equate providing a complete answer with reducing cognitive burden. Interfaces could instead use progressive disclosure~\cite{nielsen2006progressive,shneiderman1996eyes} to break an analysis into steps that users can process incrementally. 

Preserving decision agency did not preserve psychological ownership. Participants reported similar levels of decision agency across conditions, yet a sense of ownership was lower in every AI condition than in \Reader{}, with \Coevolving{} producing the lowest sense of ownership. This distinction is particularly visible in \Coevolving{}: participants could choose which questions to pursue and thereby direct the preparation process, while the AI generated much of the content that accumulated in the resulting artifact. Participants could therefore shape the direction of the work without fully authoring what it produced. Thus, preserving opportunities to choose what the AI does may be insufficient to preserve workers’ relationship to what is ultimately produced. 

This loss of the sense of ownership may also have longer-term consequences for workers. A persistent separation between workers and the products of their labor is central to research on work alienation, which is associated with poorer workplace attitudes and outcomes~\cite{nair2010alienation,shantz2015alienation}. Our 15-minute preparation task cannot establish such longer-term consequences, but it shows that improved output and preserved decision agency are not sufficient evidence that an AI-supported workflow benefits workers. To improve ownership, systems could instead preserve consequential but lightweight forms of authorship. For example, an AI might generate an adaptable scaffold while leaving users to select, revise, and author its content.

\subsection{AI Literacy Cannot Substitute for Responsible Workflow Design}

AI literacy can help people evaluate AI outputs, communicate their needs, and collaborate with AI systems~\cite{long2020ailiteracy}. Training can improve these abilities. For example, teaching people to articulate what they need from a model can improve their interactions with AI more than conventional prompt-engineering instruction~\cite{ma2025rope}. However, relying on literacy as the primary means of making AI useful places an additional burden on workers. People placed in the loop of an automated system often absorb responsibility for its failures while holding little practical control over them~\cite{elish2019crumple, green2019algorithm}. They must determine not only how to perform their professional task, but also how to organize the AI's role within it.

Domain expertise did not resolve this problem. Participants with more negotiation experience initiated chat at similar rates and asked similar types of questions as participants with less experience, even though negotiation experience predicted better preparation when participants worked without AI in \Reader{}. Knowing more about how to perform the underlying professional task therefore did not necessarily translate into knowing how to delegate that task productively to AI. These findings suggest that designers should not assume that either domain expertise or familiarity with AI will enable workers to compensate for an underspecified workflow.

Nor should AI assistance be assumed to reduce the cognitive burden of managing that workflow. None of the AI-supported conditions reduced subjective effort relative to \Reader{}, and \Prefilled{}, the condition that required participants to generate the least analysis themselves, was the most demanding of the three AI conditions. Workers may therefore face cognitive work not only in performing the task itself, but also in interpreting AI output, deciding what assistance is useful, and integrating that assistance into their own work. AI literacy may make workers better equipped to perform some of this work, but it does not make the work disappear.


Deploying AI at work therefore requires workflow design. Work design research argues that the consequences of automation for workers depend on how roles, autonomy, and task boundaries are configured around it~\cite{parker2022workdesign}, and that the division of control between algorithms and workers is an organizational choice rather than a technical consequence~\cite{kellogg2020algorithms}. Organizations should accordingly work with employees and domain experts to determine which judgments remain with workers, which activities AI supports, and how authorship, verification, decision rights, and accountability are allocated~\cite{muller1993participatory,delgado2023participatory}. Otherwise, these choices do not disappear. They are made implicitly and often left for individual workers to resolve in practice. Treating these responsibilities primarily as matters of AI literacy risks making workers responsible for compensating for shortcomings that could instead be addressed through the design of the workflow itself. 

%% file: sections/07-limitations.tex

%% file: sections/08-conclusion.tex
\section{Conclusion}
\label{sec:conclusion}

General-purpose AI does not remove the need to design how professional work is structured. In negotiation preparation, AI support improved case coverage, while workflows that made professional task structure visible produced an additional coverage gain over open-ended chat (Figure~\ref{fig:nested-outcome-comparisons}). Yet exposing the same scaffold as a completed analysis or as an incrementally developed artifact produced different forms of engagement: the co-evolving workflow elicited a broader analytic repertoire and required less subjective effort than the prefilled workflow, despite producing similar overall coverage (Figures~\ref{fig:prompt-use-patterns} and~\ref{fig:nested-outcome-comparisons}). A scaffold therefore shapes work not only through what it represents, but through how people encounter and develop the analysis it organizes.

Professional AI systems should neither require users to reconstruct a domain workflow through prompting nor assume that presenting a comprehensive answer provides effective support. Designers can instead make professional task structure visible through persistent, inspectable, and revisable artifacts, while deliberately allocating what users direct, author, verify, and retain. The central design object is thus not the prompt or the generated output alone, but the workflow through which people and AI build an analysis together. Designing that workflow makes it possible to pursue preparation quality, productive analytic engagement, manageable effort, and a meaningful relationship to the resulting work as distinct design goals.

%% file: sections/09-appendix.tex
\appendix

\section{Case Materials}
\label{app:case-files}

Participants saw the 15 documents in Table~\ref{tab:case-files}, totaling 6{,}977 words. Table~\ref{tab:case-files} describes the corpus at the document level.

\begin{table}[h]
\centering
\caption{The 15 case documents visible to participants, with length in words.}
\label{tab:case-files}
\small
\begin{tabular}{lr}
\toprule
Document                                          & Words \\
\midrule
FWB role card                                     & 935 \\
UN logistics coordination note                    & 596 \\
Refugee committee message                         & 574 \\
Donor compliance exception Q\&A                   & 565 \\
Operations and logistics email                    & 561 \\
Market price and labor bulletin                   & 548 \\
Host community council petition                   & 508 \\
Minutes: FWB field officer with the Commander     & 488 \\
Health for All access update                      & 485 \\
Minutes: the Commander with the FWB field officer & 407 \\
FWB assisting refugees in Alpha                   & 391 \\
Message from the FWB country director             & 283 \\
Encrypted note from a local contact               & 281 \\
WhatsApp message from the FWB field officer       & 191 \\
Media monitoring, local news excerpt              & 164 \\
\midrule
Total                                             & 6{,}977 \\
\bottomrule
\end{tabular}
\end{table}

The corpus combined role-defining documents, operational records, stakeholder claims, and secondary reports. Participants were not told which documents to privilege; evaluating provenance and resolving conflicting claims were part of the preparation task.

\section{System and Interface Details}
\label{app:system}

\subsection{Common Task Shell}

All four interfaces used the same browser application, case corpus, document order, preparation timer, and note-taking area. During preparation, participants selected documents from a numbered file list and read them in a central pane. When preparation ended, the complete case files were removed. Participants retained their notes, chat transcripts where available, and structured artifacts in the scaffolded conditions while answering the written questions.

\subsection{Interface Tutorial}
\label{app:tutorial}

Before the timed preparation phase, participants completed a non-skippable tutorial using a neutral community-center scheduling case unrelated to the experimental scenario. The tutorial reproduced the participant's assigned interface using predefined practice content and made no live AI calls. All versions introduced the document list, case reader, and condition-specific workspace. The \Reader{} tutorial introduced the notes panel; the \Chatbot{} tutorial introduced the conversational assistant and passage-to-AI interaction; the \Prefilled{} tutorial introduced the completed, cited preparation scaffold and conversational assistant; and the \Coevolving{} tutorial introduced the initially empty scaffold, conversational assistant, and artifact-update workflow. After the tutorial, all practice content was cleared and the experimental case was loaded. The \Tprep-minute preparation timer began only after this process was complete.
\paragraph{Practice documents.} The tutorial used the following two documents:

\begin{quote}
\small
\textbf{Practice case---The Saturday room clash.} This is a practice case for the tutorial. It is not the case you will negotiate.

The Riverside Community Center has one large hall. For years the art club has used it on Saturday mornings, the only time the building's accessible entrance is staffed. The art club cannot afford to lose its only accessible morning slot. This spring a new youth coding club received a small grant and asked for the same hall. The coding club's grant requires at least one weekend session each month, or the funding is withdrawn.

The two clubs meet next week to work out a schedule. The center's director has said she will approve whatever the clubs agree on, as long as neither group is pushed out of the building.
\end{quote}

\begin{quote}
\small
\textbf{Practice note---From the facilities manager.} This is a practice document for the tutorial.

Quick note: the smaller annex room is free most Saturdays, but it has no projector. The hall calendar shows two Saturdays a month with no bookings after 1pm. If a group wants storage space for supplies, the basement cage is available on request.
\end{quote}

\paragraph{Tutorial popovers.} The following reproduces the rendered tutorial text. Text shared across conditions appeared as follows:

\begin{description}
\item[Quick interface overview.] This short tutorial uses a \textbf{practice example}---a community-center scheduling clash, \emph{not} the case you will negotiate. It only shows where things are; the real case and the 15-minute timer start right after.
\item[Left: case documents.] The left side is the document list. Use the numbered files to switch between case documents during the reading stage.
\item[Practice done.] Start the \textbf{15-minute paid task}. The practice content will clear, the real case will load, and the reading timer will begin. The final button read \textit{Start the 15-minute task}.
\end{description}

The condition-specific popovers appeared between these shared opening and closing prompts:

\begin{description}
\item[\Reader.] \textbf{Middle: reading area.} The middle is where the selected document opens. Read the files here and keep important points in your Notes panel. \textbf{Right: notes.} The right side is your private scratchpad. Save the points you will want when answering the written questions later.

\item[\Chatbot.] \textbf{Middle: reading area.} The middle is where the selected document opens. Highlight a passage and choose \textbf{Copy to AI} when you want to ask the assistant about it. \textbf{Right: notes.} The right side is your private scratchpad. Save the points you will want when answering the written questions later. \textbf{Below the reader: AI assistant.} The strip under the document is an AI assistant. Ask it where to start or what a passage means---the conversation opens right there, under what you are reading.

\item[\Prefilled.] \textbf{Middle: reading area.} The middle is where the selected document opens. Read for issues, priorities, objectives, redlines, interests, possible proposals, and things to avoid. \textbf{Right: your preparation sheet.} Three modules the AI has already filled from the case files: what each side wants, where they agree and clash, and how a deal could work. Every point \textbf{cites its source}---click a citation chip to check it against the document. \textbf{Below the reader: AI assistant.} The strip under the document is an AI assistant. Ask it where to start or what a passage means---the conversation opens right there, under what you are reading.

\item[\Coevolving.] \textbf{Middle: reading area.} The middle is where the selected document opens. Read for issues, priorities, objectives, redlines, interests, possible proposals, and things to avoid. \textbf{Right: your preparation sheet.} It starts empty. Your first conversation with the chatbot fills the sheet automatically. Later changes appear as focused revisions inside the affected cells; apply the update or keep the current sheet. \textbf{Below the reader: AI assistant.} The strip under the document is an AI assistant. Ask it where to start or what a passage means---the conversation opens right there, under what you are reading.
\end{description}

\subsection{Conversational Support}

The \Chatbot assistant received the assigned side (FWB), all case documents, and the recent conversation. Its prompt required answers to be grounded in the supplied documents, to distinguish direct records from rumors and other mixed-signal material, and to state when the documents did not support an answer. Participants could type a question or send a selected passage from the reader. The \Chatbot{}, \Prefilled and \Coevolving assistant used the same document-grounded question-answering format while \Coevolving's transcript also served as input to the structured artifact.

\subsection{Structured Preparation Artifacts}

Both scaffold conditions used the same three modules. \textit{Sides and needs} recorded each party's stated positions and what lay underneath them. \textit{Agreement and conflict} recorded shared ground and core tensions. \textit{Deal options} recorded workable packages, red lines and bottom lines, and actors or channels that could support implementation. Each generated claim could carry up to two source passages. Selecting a citation opened the source document at the quoted passage. A separate evidence-highlighting pass marked case-file passages relevant to the active scaffold section; participants could save a highlighted passage to their notes with one click.

In \Prefilled, the system generated all three modules from the case corpus before the participant began. This module is cached across participants. The cache key included the model, prompt version, assigned role, and file contents. Because all participants in the experiment had the same role and case files, the cached initial artifact was identical across participants. Participant-requested revisions were generated live and were not cached.

In \Coevolving, the artifact began empty. At each conversation turn, an LLM summarizes the conversation into the relevant cells in the scaffold. A separate pass attached supporting case passages without changing the generated text. Content entering an empty cell appeared immediately. A proposed change to an already populated cell was shown as a revision that the participant could apply or reject. The \Coevolving interface also marked one area that the conversation had not yet developed at each conversation turn. 

\subsection{Models and Prompt Constraints}

Table~\ref{tab:system-models} reports the model used by each participant-facing system component. Chat requests returned text. Artifact generation, extraction, citation, and evidence scans used schema-constrained JSON. 

\begin{table}[t]
\centering
\caption{Models and prompt constraints for participant-facing system components.}
\label{tab:system-models}
\small
\begin{tabular}{p{0.16\textwidth}p{0.16\textwidth}p{0.16\textwidth}p{0.38\textwidth}}
\toprule
Component & Condition & Model & Prompt constraint \\
\midrule
Open-ended assistant & \Chatbot, \Coevolving, \Prefilled{} & \texttt{gpt-5.6-luna}, no reasoning effort & Answer from the case documents and acknowledge unsupported answers \\
Initial artifact and requested revisions & \Prefilled & \texttt{gpt-5.6-terra} & Produce one schema-constrained module grounded in cited case passages \\
Conversation-to-artifact update & \Coevolving & \texttt{gpt-5.6-terra}, no reasoning effort & Make the smallest cell-level changes supported by explicit conversation content \\
Evidence scan & \Coevolving{}, \Prefilled{} & \texttt{gpt-5.4-mini} & Identify potentially relevant passages without deciding what the participant should conclude \\
\bottomrule
\end{tabular}
\end{table}

\section{Written Preparation Questions}
\label{app:questions}

Before the assessment began, participants saw the following instruction.

\begin{quote}
You do not need to write detailed prose. Concise bullet points and incomplete sentences are enough and will not be penalized.
\end{quote}

The questions appeared in four timed sections. A participant could continue early and carry the unused time into the next section but could not return to an earlier section.

\paragraph{Section 1, Your side, 5 minutes}
\begin{enumerate}
    \item What issues are being negotiated? Rank them by priority for your side.
    \item What are Food Without Borders' objectives?
    \item What are Food Without Borders' redlines or non-negotiables?
\end{enumerate}

\paragraph{Section 2, Opposing side, 3.5 minutes}
\begin{enumerate}
    \setcounter{enumi}{3}
    \item What are the Camp Commander's objectives?
    \item What are the Camp Commander's redlines or non-negotiables?
\end{enumerate}

\paragraph{Section 3, Overlap and conflict, 3.5 minutes}
\begin{enumerate}
    \setcounter{enumi}{5}
    \item What common ground do the two sides have? Include both points they have explicitly accepted and underlying goals or needs they share.
    \item Where are the main conflicts or disagreements between the two sides?
\end{enumerate}

\paragraph{Section 4, Scenario planning, 5 minutes}
\begin{enumerate}
    \setcounter{enumi}{7}
    \item What realistic package would you propose?
    \item What risks could prevent your proposed package from working?
    \item What would you ask to clarify before or during the negotiation?
\end{enumerate}

\section{Survey Instruments}
\label{app:surveys}

\subsection{Demographic and Background Items}

Table~\ref{tab:intake-items} reproduces the intake questions and response formats. Prior participation, age, AI-tool use, English proficiency, education, negotiation experience, and the four thinking-style items were required. Gender, computer use, and both country fields were optional.

\begin{table}[t]
\centering
\caption{Demographic and background questions shown before the task.}
\label{tab:intake-items}
\footnotesize
\begin{tabular}{p{0.42\textwidth}p{0.51\textwidth}}
\toprule
Item & Response options or format \\
\midrule
Have you taken this test before? & No; Yes \\
How old are you? & Numeric entry from 18 to 100 \\
What is your gender? & Male; Female; Non-binary; Prefer to self-describe; Prefer not to say \\
How often do you use a computer? & Once a week or less; A few times a week; A couple of hours most days; Many hours on most days \\
How often do you use ChatGPT or similar AI tools? & Never; Less than once a month; A few times a month; A few times a week; Daily or almost daily \\
How would you describe your English proficiency? & Native or near-native; Advanced; Intermediate; Basic; Prefer not to say \\
Highest education level completed or currently pursuing & Pre-high school; High school or equivalent; Some college; Bachelor's degree; Master's or professional degree; PhD, doctorate, or professional doctorate; Prefer not to say \\
In which country did you spend most of your childhood? & Free-text country entry \\
In which country do you currently live? & Free-text country entry \\
Negotiation experience & No formal negotiation experience; Some coursework or training; Negotiation experience through work or volunteering; Frequent negotiation as part of my role \\
\bottomrule
\end{tabular}
\end{table}

\subsection{Need for Cognition Items}

Participants rated each item from 1, \textit{strongly disagree}, to 5, \textit{strongly agree}.

\begin{enumerate}
    \item I would prefer complex to simple problems.
    \item I find satisfaction in deliberating hard and for long hours.
    \item Thinking is not my idea of fun.
    \item I would rather do something that requires little thought than something that is sure to challenge my thinking abilities.
\end{enumerate}

The third and fourth items were reverse-coded before the four responses were averaged.

\subsection{Mid-study Items}

Immediately after preparation, participants rated the following items from 1, \textit{strongly disagree}, to 7, \textit{strongly agree}.

\begin{enumerate}
    \item This preparation task required a lot of mental effort.
    \item I am confident that I prepared effectively for this negotiation.
    \item I am confident in my ability to walk into this negotiation with a clear strategy.
\end{enumerate}

\subsection{Post-study Items}

The post-study survey first asked the following binary item.

\begin{quote}
Before this study, had you seen or heard of CCHN or other negotiation frameworks/tools, such as Island of Agreements, Iceberg, Redline/Bottom Line, BATNA/ZOPA, or interest-based negotiation?
\end{quote}

The response options were \textit{yes} and \textit{no}. Participants then rated the following agreement items from 1, \textit{strongly disagree}, to 7, \textit{strongly agree}.

\paragraph{Decision agency}
\begin{enumerate}
    \item I had control over how I prepared my negotiation answers.
    \item I could decide what information or suggestions to use, revise, or ignore.
    \item The system constrained me in ways that limited my control over the preparation.
\end{enumerate}

The third item was reverse-coded before aggregation.

\paragraph{Psychological ownership}
\begin{enumerate}
    \item The preparation I produced felt like my own work.
    \item I felt responsible for the content of the preparation I produced.
    \item The final preparation reflected my own reasoning.
\end{enumerate}

\paragraph{Case understanding and usefulness}
\begin{enumerate}
    \item The preparation process helped me understand the negotiation case.
    \item The preparation I produced would be useful for entering the negotiation.
\end{enumerate}

Participants rated two final items from 1, \textit{very low}, to 7, \textit{very high}.

\paragraph{Mental effort and frustration}
\begin{enumerate}
    \item How hard did you have to work mentally to answer the written questions with the information available to you?
    \item How frustrating was it to use the available information to prepare your answers?
\end{enumerate}

\section{Preparation-Coverage Measure and Validation}
\label{app:rubric}

The reported preparation-performance measure covers Questions 1--9 (Table~\ref{tab:rubric-structure}). Questions 1--7 and 9 use atomic, case-grounded criteria evaluated only against their assigned answer. Question 8 uses seven package-issue dimensions and direct option coding. Overall preparation coverage is the unweighted mean of the nine question-level scores, so questions with more criteria do not receive greater weight. Question 10 remained an open-ended elicitation item rather than a quantitatively scored dimension.

\begin{table}[t]
\centering
\caption{Structure of the reported negotiation-preparation measure. The professional scaffold represented the first eight dimensions but did not explicitly represent package risks.}
\label{tab:rubric-structure}
\small
\begin{tabular*}{\columnwidth}{@{\extracolsep{\fill}}rlc@{}}
\toprule
Question & Preparation dimension & Represented in scaffold \\
\midrule
1 & Negotiated issues and priorities & Yes \\
2 & FWB objectives & Yes \\
3 & FWB redlines & Yes \\
4 & Counterpart objectives & Yes \\
5 & Counterpart redlines & Yes \\
6 & Common ground & Yes \\
7 & Main conflicts & Yes \\
8 & Realistic agreement package & Yes \\
9 & Package risks & No \\
\bottomrule
\end{tabular*}
\end{table}

\begin{table}[t]
\centering
\caption{Independent expert agreement before adjudication and held-out performance of the frozen \texttt{gpt-5.6-sol} grading programs. Expert agreement covers Questions 1--9 for 100 participants. Overall coverage reliability uses a two-way, absolute-agreement, single-measure ICC for the experts' nine-question mean scores. The participant-disjoint test split evaluates a binary-criteria grader for Questions 1--7 and 9 and a separate package-issue grader for Question 8.}
\label{tab:grading-validation}
\small
\begin{tabular*}{\columnwidth}{@{\extracolsep{\fill}}llr@{}}
\toprule
Evaluation & Evaluation sample & Value \\
\midrule
\multicolumn{3}{l}{\textit{Independent expert coding before adjudication}} \\
Pooled item-level Cohen's $\kappa$ & 100 participants $\times$ 61 items & .730 \\
Overall coverage ICC(A,1) & 100 participants $\times$ 9 questions & .949 \\
Mean absolute coverage difference & 100 participants $\times$ 9 questions & .042 \\
\midrule
\multicolumn{3}{l}{\textit{Frozen binary-criteria grader: Questions 1--7 and 9}} \\
Binary-criteria macro-F1 & 20 participants $\times$ 54 items & .758 \\
Binary-criteria micro-F1 & 20 participants $\times$ 54 items & .821 \\
Exact criterion-set accuracy & 20 participants $\times$ 8 questions & .463 \\
\midrule
\multicolumn{3}{l}{\textit{Frozen package-issue grader: Question 8}} \\
Question 8 issue-coverage accuracy & 20 participants $\times$ 7 issues & .921 \\
\bottomrule
\end{tabular*}
\end{table}

\section{Prompt Code Frame}
\label{app:code-frame}

Each participant turn received exactly one request-form code and at least one purpose code. Request form describes how the turn entered the conversation: \textit{authored request}, \textit{passage plus request}, or \textit{passage only}. Purpose describes why the participant recruited AI (Table~\ref{tab:code-frame}). Purpose codes were multilabel except that \textit{other} was exclusive and a passage-only turn received only \textit{reading support}. Coders used preceding conversation only to resolve references and short follow-ups; they did not infer a purpose from the assistant's response.

\begin{table}[t]
\centering
\caption{Request-form code frame. Definitions describe interface actions rather than reproducing participant text.}
\label{tab:request-form-frame}
\small
\begin{tabular}{p{0.25\columnwidth}p{0.65\columnwidth}}
\toprule
Form & Definition \\
\midrule
Authored request & Participant entered a message without attaching a selected passage \\
Passage plus request & Participant attached a selected passage and entered an additional request \\
Passage only & Participant sent a selected passage without additional authored text \\
\bottomrule
\end{tabular}
\end{table}

Own-side analysis, counterpart analysis, cross-party synthesis, and strategy or package development form the four-purpose analytic-breadth measure. Merely naming a party did not qualify as analysis: the turn had to ask for interpretation, evaluation, or inference. A turn received both cross-party synthesis and strategy or package when it asked AI both to integrate the parties' positions and recommend an action. Roleplay required addressing AI as the counterpart, rather than asking about the counterpart in the third person.

\begin{table*}[t]
\centering
\caption{The 12-purpose prompt code frame. Examples are lightly paraphrased from representative participant turns to protect participant text.}
\label{tab:code-frame}
\footnotesize
\begin{tabular*}{\textwidth}{@{\extracolsep{\fill}}p{0.18\textwidth}p{0.39\textwidth}p{0.33\textwidth}@{}}
\toprule
Category & Definition & Representative example \\
\midrule
Reading support & Processes reading material by summarizing, explaining, simplifying, or restating it; also includes passage-only turns & ``Summarize this passage in plain language.'' \\
Fact retrieval & Retrieves a specific fact directly available in the case record & ``What access restrictions did the Commander impose?'' \\
Own-side analysis & Interprets FWB's objectives, interests, constraints, redlines, authority, or risks & ``Which constraints are non-negotiable for FWB?'' \\
Counterparty analysis & Interprets the Commander's objectives, concerns, constraints, likely reactions, or risks & ``Why might the Commander resist an independent assessment?'' \\
Cross-party synthesis & Compares or integrates the parties' positions, evidence, interests, constraints, or disputed facts & ``Where do the two sides' positions overlap and conflict?'' \\
Strategy or package & Recommends negotiation actions or constructs terms, trades, sequencing, openings, or packages & ``Propose a package that protects FWB's redlines while addressing the Commander's concerns.'' \\
Deliverable drafting & Produces answer-ready text for the participant's written deliverable & ``Draft a concise answer describing the proposed package.'' \\
Roleplay & Treats the assistant as the counterpart or requests a simulated negotiation & ``Act as the Commander and respond to my opening proposal.'' \\
Verification or challenge & Checks, defends, corrects, or requests evidence for a claim or prior output & ``Is that claim supported by the case files?'' \\
Interaction control & Controls continuation or presentation without adding a new substantive purpose & ``Continue, but make it shorter.'' \\
Task meta & Asks about the study task, interface, available tools, timing, or submission procedure & ``How much time do I have left?'' \\
Other & Expresses an identifiable purpose not represented above & ``Thank you---that helps.'' \\
\bottomrule
\end{tabular*}
\end{table*}

\begin{table}[t]
\centering
\caption{Human agreement and held-out classifier performance for prompt coding. Human purpose agreement was calculated on the 13 initial labels before the two reading-related labels were merged. Classifier purpose metrics use the final 12-purpose frame. Macro-F1 averages purposes represented in the held-out set.}
\label{tab:prompt-classifier-validation}
\small
\begin{tabular*}{\columnwidth}{@{\extracolsep{\fill}}lrr@{}}
\toprule
Evaluation & $n$ & Value \\
\midrule
\multicolumn{3}{l}{\textit{Independent human coding before adjudication}} \\
Request-form Cohen's $\kappa$ & 400 & .991 \\
Macro binary $\kappa$ across purposes & 400 & .853 \\
Exact purpose-set agreement & 400 & .815 \\
Mean purpose-set Jaccard agreement & 400 & .871 \\
\midrule
\multicolumn{3}{l}{\textit{Frozen classifier on held-out messages}} \\
Request-form accuracy & 80 & .975 \\
Purpose micro-F1 & 80 & .856 \\
Purpose macro-F1 & 80 & .813 \\
Exact purpose-set accuracy & 80 & .788 \\
\bottomrule
\end{tabular*}
\end{table}

\begin{table}[t]
\centering
\caption{Before-adjudication human agreement by purpose on 400 independently double-coded turns. Cohen's $\kappa$ is undefined for roleplay because neither coder assigned a positive label. The first two labels were subsequently merged as reading support.}
\label{tab:prompt-purpose-irr}
\small
\begin{tabular*}{\columnwidth}{@{\extracolsep{\fill}}lr@{}}
\toprule
Initial purpose label & Cohen's $\kappa$ \\
\midrule
Passage-only offload & 1.000 \\
Material explanation & .912 \\
Fact retrieval & .839 \\
Own-side analysis & .799 \\
Counterparty analysis & .885 \\
Cross-party synthesis & .841 \\
Strategy or package & .854 \\
Deliverable drafting & .773 \\
Roleplay & --- \\
Verification or challenge & .877 \\
Interaction control & .890 \\
Task meta & .859 \\
Other & .712 \\
\bottomrule
\end{tabular*}
\end{table}

\begin{table}[t]
\centering
\caption{Development-set selection of the prompt-purpose classifier. Panel A compares DSPy programs using \texttt{gpt-5.4-mini}; the prespecified selection prioritized macro-F1, subject to request-form accuracy and verification-recall guards. Panel B compares candidate models using the retained zero-shot program. Both panels use the same condition-balanced development set of 80 turns and the initial 13-label purpose frame; the two reading-related labels were merged deterministically after classification.}
\label{tab:prompt-program-selection}
\small
\begin{tabular*}{\columnwidth}{@{\extracolsep{\fill}}lrrrr@{}}
\toprule
Program or model & Request-form accuracy & Macro-F1 & Micro-F1 & Exact purpose set \\
\midrule
\multicolumn{5}{l}{\textit{Panel A: DSPy program comparison}} \\
Zero-shot & 1.000 & \textbf{.521} & .610 & \textbf{.538} \\
Four-shot & .938 & .416 & .505 & .325 \\
Eight-shot & .963 & .459 & .554 & .288 \\
Twelve-shot & .963 & .429 & .513 & .263 \\
Sixteen-shot & .950 & .418 & .524 & .288 \\
GEPA-optimized & 1.000 & .503 & \textbf{.635} & .525 \\
\midrule
\multicolumn{5}{l}{\textit{Panel B: model comparison with the retained zero-shot program}} \\
\texttt{gpt-5.4-mini} & .988 & .564 & .634 & .538 \\
\texttt{gpt-5.6-luna} & .988 & \textbf{.569} & \textbf{.698} & \textbf{.638} \\
\texttt{gpt-5.6-terra} & \textbf{1.000} & .502 & .690 & \textbf{.638} \\
\bottomrule
\end{tabular*}
\end{table}

\begin{figure*}[t]
\centering
\includegraphics[width=\textwidth]{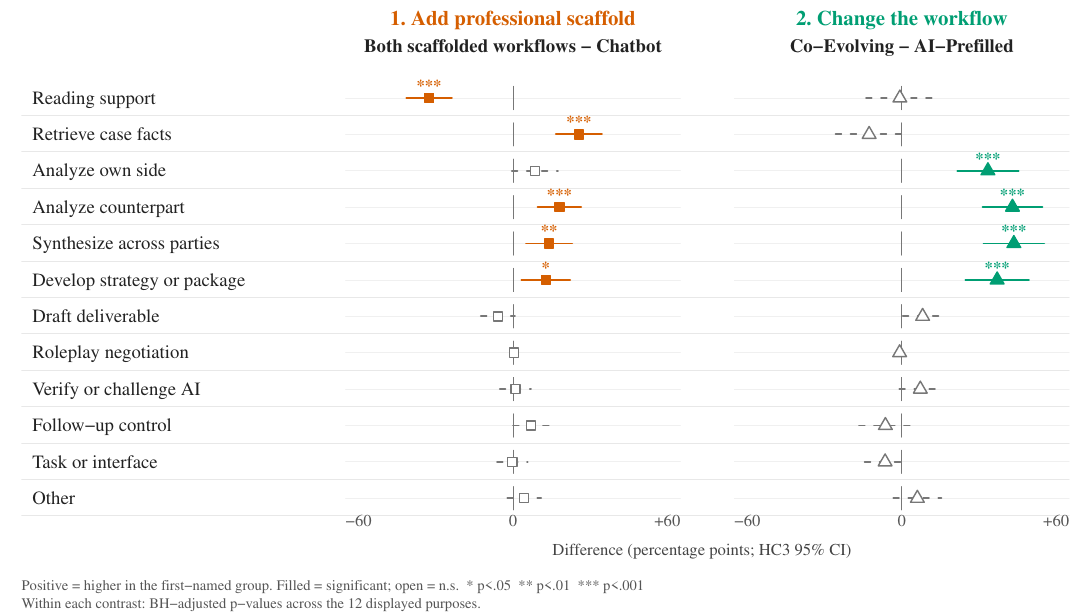}
\caption{\textbf{The largest interface-related shifts concerned reading support and the four analytic purposes.} This complete inventory reports participant-level incidence of all 12 prompt purposes. Points are percentage-point differences with HC3 95\% confidence intervals. Stars use Benjamini--Hochberg correction across the 12 purposes within each contrast. Filled colored markers, solid intervals, and stars indicate adjusted $p<.05$; open gray markers and dashed intervals are nonsignificant.}
\Description{Two forest plots compare both scaffolded workflows with \Chatbot{} and \Coevolving{} with \Prefilled{} across all 12 prompt purposes. The strongest differences concern reading support and the four analytic purposes. Several less common purposes show no detectable differences.}
\label{fig:prompt-purpose-inventory}
\end{figure*}

\section{Analytic Model Configuration}
\label{app:prompts}

Table~\ref{tab:analytic-models} distinguishes the models used for measurement from the participant-facing models in Appendix Table~\ref{tab:system-models}. All analytic calls used schema-constrained outputs. The preparation-quality judge received one written response and only the frozen criteria for that question. The prompt-purpose classifier received one target turn and up to four preceding turns, which it could use only to resolve context.

\begin{table}[t]
\centering
\caption{Models used in the analytic pipeline.}
\label{tab:analytic-models}
\small
\begin{tabular}{p{0.27\columnwidth}p{0.23\columnwidth}p{0.40\columnwidth}}
\toprule
Component & Model & Output \\
\midrule
Preparation-quality judge & \texttt{gpt-5.6-sol}, no reasoning effort & Criterion-level support judgments for Questions 1--7 and 9 \\
Package option coder & \texttt{gpt-5.6-sol}, no reasoning effort & Question 8 mappings to the frozen 29-option table \\
Prompt-purpose classifier & \texttt{gpt-5.6-luna} & Request form and multilabel purpose codes \\
Solution-diversity embedding & \texttt{text-embedding-3-large} & One normalized embedding per non-empty package response \\
\bottomrule
\end{tabular}
\end{table}

\section{Exact Results and Additional Analyses}
\label{app:additional}

\begin{figure*}[t]
\centering
\includegraphics[width=\textwidth]{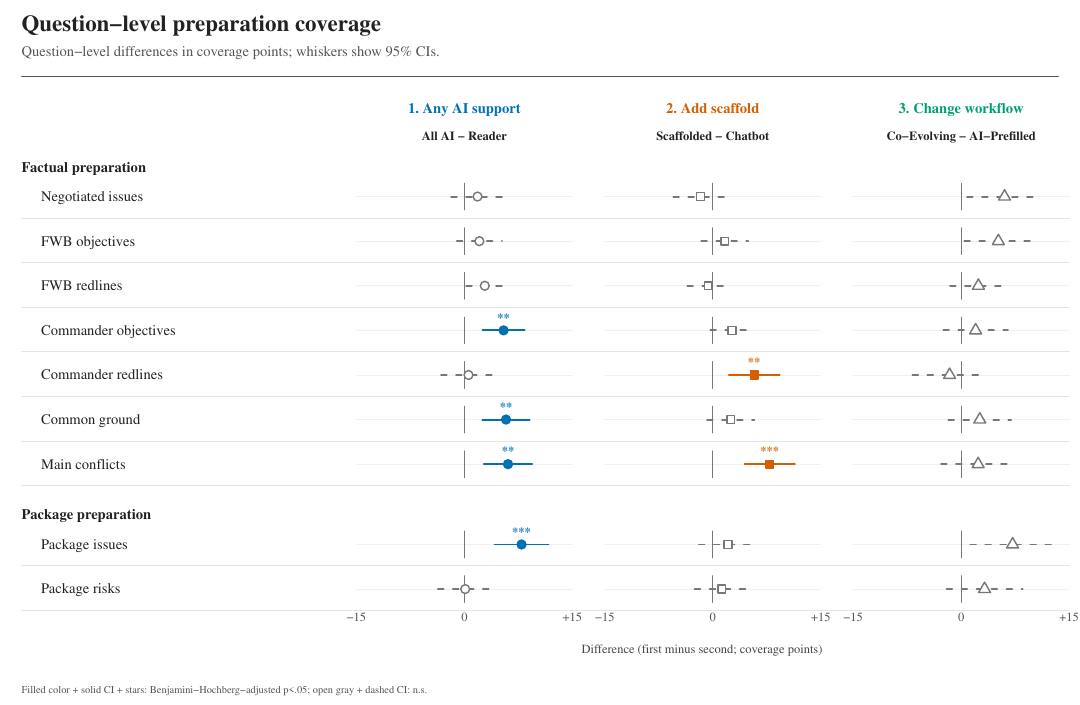}
\caption{\textbf{AI-related coverage gains appeared in both factual analysis and package construction.} Points are nested differences in question-level coverage with 95\% confidence intervals. Stars use Benjamini--Hochberg correction across the nine questions within each comparison. Filled colored markers and solid intervals indicate adjusted $p<.05$; open gray markers and dashed intervals are nonsignificant.}
\Description{Three aligned forest plots show seven factual-preparation questions and two package-preparation questions. AI support improves the Commander's objectives, common ground, main conflicts, and package-issue coverage. Scaffolding improves the Commander's redlines and main conflicts. No Co-Evolving versus AI-Prefilled comparison remains significant after correction.}
\label{fig:question-quality-comparisons}
\end{figure*}

\begin{table*}[t]
\centering
\caption{Overall outcomes by condition and nested contrast. Condition columns report means with standard deviations in parentheses; contrast columns report Hedges' $g$. Package-quality significance uses Benjamini--Hochberg correction within contrast; experience outcomes use Holm correction within outcome. Positive values favor the first-named group. Bold cells mark detectable comparisons. $^{*}p<.05$, $^{**}p<.01$, $^{***}p<.001$.}
\label{tab:overall-outcomes}
\scriptsize
\setlength{\tabcolsep}{2.5pt}
\begin{tabular*}{\textwidth}{@{\extracolsep{\fill}}lrrrrrrr@{}}
\toprule
& \multicolumn{4}{c}{Condition summary, $M$ (SD)} & \multicolumn{3}{c}{Nested contrast, Hedges' $g$} \\
\cmidrule(lr){2-5}\cmidrule(lr){6-8}
Outcome & \Reader{} & \Chatbot{} & \Prefilled{} & \Coevolving{} & \shortstack{Any AI\\$-$ \Reader{}} & \shortstack{Scaffolded\\$-$ \Chatbot{}} & \shortstack{\Coevolving{}\\$-$ \Prefilled{}} \\
\midrule
Overall preparation coverage & 26.54 (14.40) & 28.82 (12.88) & 29.82 (14.28) & 33.01 (16.91) & \textbf{$0.25^{**}$} & \textbf{$0.16^{*}$} & $0.21$ \\
Risk coverage & 34.49 (21.91) & 33.87 (22.14) & 33.68 (22.36) & 36.93 (26.49) & $0.00$ & $0.05$ & $0.13$ \\
Risk relevance & 88.31 (26.45) & 86.28 (26.92) & 86.82 (27.62) & 87.59 (24.59) & $-0.06$ & $0.03$ & $0.03$ \\
Preparation effort & 6.39 (0.96) & 6.35 (1.03) & 6.55 (0.92) & 6.17 (1.30) & $-0.02$ & 0.04 & \textbf{$-0.35^{**}$} \\
Answer-writing effort & 5.98 (1.14) & 5.82 (1.31) & 5.89 (1.31) & 5.73 (1.37) & $-0.12$ & 0.00 & $-0.12$ \\
Frustration & 4.56 (1.94) & 4.61 (1.92) & 4.58 (2.08) & 4.48 (1.99) & 0.00 & $-0.04$ & $-0.05$ \\
Decision agency & 5.04 (1.35) & 5.11 (1.32) & 5.12 (1.22) & 5.11 (1.27) & 0.06 & 0.01 & $-0.01$ \\
Psychological ownership & 5.91 (1.10) & 5.52 (1.38) & 5.23 (1.52) & 5.10 (1.60) & \textbf{$-0.42^{***}$} & \textbf{$-0.23^{*}$} & $-0.08$ \\
\bottomrule
\end{tabular*}
\end{table*}

\begin{table*}[t]
\centering
\caption{Question-level coverage by condition and nested contrast. Condition columns report mean coverage from 0 to 100; contrast columns report differences in coverage points. Stars use Benjamini--Hochberg correction across nine questions within each contrast.}
\label{tab:per-question}
\small
\setlength{\tabcolsep}{4pt}
\begin{tabular*}{\textwidth}{@{\extracolsep{\fill}}lrrrrrrr@{}}
\toprule
& \multicolumn{4}{c}{Mean coverage} & \multicolumn{3}{c}{Nested difference} \\
\cmidrule(lr){2-5}\cmidrule(lr){6-8}
Question & \Reader{} & \Chatbot{} & \Prefilled{} & \Coevolving{} & \shortstack{Any AI\\$-$ \Reader{}} & \shortstack{Scaffolded\\$-$ \Chatbot{}} & \shortstack{\Coevolving{}\\$-$ \Prefilled{}} \\
\midrule
\multicolumn{8}{l}{\textit{Factual preparation}} \\
Negotiated issues & 41.8 & 44.6 & 40.3 & 46.3 & $+1.8$ & $-1.7$ & $+6.0$ \\
FWB objectives & 24.2 & 25.3 & 24.8 & 29.9 & $+2.1$ & $+1.7$ & $+5.2$ \\
FWB redlines & 15.3 & 18.5 & 16.8 & 19.2 & $+2.8$ & $-0.6$ & $+2.4$ \\
Commander objectives & 27.3 & 31.2 & 33.0 & 35.0 & \textbf{$+5.4^{**}$} & $+2.7$ & $+2.0$ \\
Commander redlines & 15.7 & 13.0 & 19.4 & 17.8 & $+0.6$ & \textbf{$+5.8^{**}$} & $-1.6$ \\
Common ground & 20.7 & 25.1 & 26.3 & 28.9 & \textbf{$+5.8^{**}$} & $+2.5$ & $+2.6$ \\
Main conflicts & 27.0 & 28.5 & 35.4 & 37.7 & \textbf{$+6.1^{**}$} & \textbf{$+7.9^{***}$} & $+2.4$ \\
\midrule
\multicolumn{8}{l}{\textit{Package preparation}} \\
Package issues & 32.4 & 39.1 & 38.1 & 45.3 & \textbf{$+7.9^{***}$} & $+2.1$ & $+7.1$ \\
Package risks & 34.5 & 33.9 & 33.7 & 36.9 & $+0.1$ & $+1.3$ & $+3.2$ \\
\bottomrule
\end{tabular*}
\end{table*}

\begin{table*}[t]
\centering
\caption{AI use by condition. Assistant-use rates use all participants in the three AI conditions. All other rows use participants who interacted with the assistant at least once. Purpose rows report participant-level incidence; request-form rows report the mean within-participant percentage of turns.}
\label{tab:prompt-intents}
\small
\begin{tabular*}{\textwidth}{@{\extracolsep{\fill}}lrrr@{}}
\toprule
Measure & \Chatbot{} & \Prefilled{} & \Coevolving{} \\
\midrule
Used the AI assistant at least once & 81.9\% & 62.6\% & 90.8\% \\
Participants who used the assistant, $n$ & 213 & 129 & 139 \\
Participant turns per assistant user, mean & 6.34 & 4.32 & 7.81 \\
\midrule
\multicolumn{4}{l}{\textit{Purpose incidence among assistant users}} \\
\quad Reading support & 79.3\% & 49.6\% & 48.9\% \\
\quad Fact retrieval & 23.0\% & 52.7\% & 41.0\% \\
\quad Analyze own side & 29.1\% & 20.9\% & 51.8\% \\
\quad Analyze counterpart & 22.5\% & 18.6\% & 58.3\% \\
\quad Synthesize across parties & 29.6\% & 21.7\% & 61.9\% \\
\quad Develop strategy or package & 40.8\% & 34.9\% & 69.1\% \\
\quad Deliverable drafting & 15.5\% & 6.2\% & 13.7\% \\
\quad Roleplay & 0.0\% & 0.8\% & 0.0\% \\
\quad Verification or challenge & 10.3\% & 7.8\% & 14.4\% \\
\quad Interaction control & 12.2\% & 21.7\% & 15.8\% \\
\quad Task meta & 10.3\% & 13.2\% & 7.2\% \\
\quad Other & 11.3\% & 12.4\% & 18.0\% \\
\midrule
\multicolumn{4}{l}{\textit{Analytic breadth}} \\
\quad Distinct analytic purposes, mean & 1.22 & 0.96 & 2.41 \\
\quad At least 2 of 4 purposes & 35.7\% & 23.3\% & 75.5\% \\
\quad At least 3 of 4 purposes & 20.7\% & 9.3\% & 52.5\% \\
\quad All 4 purposes & 9.9\% & 3.9\% & 27.3\% \\
\quad Expected purposes in three turns & 0.91 & 0.88 & 1.68 \\
\midrule
\multicolumn{4}{l}{\textit{Request form, mean share of participant turns}} \\
\quad Authored request & 62.9\% & 90.2\% & 86.3\% \\
\quad Passage plus authored request & 6.0\% & 1.1\% & 1.0\% \\
\quad Passage only & 31.0\% & 8.7\% & 12.7\% \\
\bottomrule
\end{tabular*}
\end{table*}

\subsection{Proposed-Package Diversity}
\label{sec:results-diversity}

\begin{figure}[t]
\centering
\includegraphics[width=0.58\textwidth]{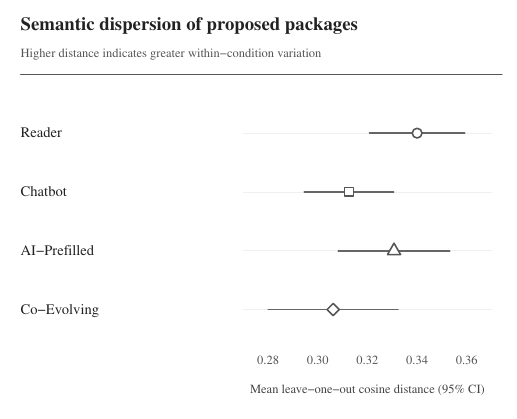}
\caption{\textbf{No condition contrast detectably changed proposed-package diversity.} Points show mean leave-one-out cosine distance with 95\% confidence intervals; exact randomization tests appear in Table~\ref{tab:solution-diversity}.}
\Description{\Reader{} has the highest mean semantic dispersion, followed by \Prefilled{}, \Chatbot{}, and \Coevolving{}, but the confidence intervals overlap.}
\label{fig:solution-diversity}
\end{figure}

AI-supported preparation did not produce a detectable change in the semantic diversity of proposed packages. Of 800 responses, 796 were non-empty and entered this exploratory analysis. Mean leave-one-out cosine distance was 0.340 in \Reader{}, 0.313 in \Chatbot{}, 0.331 in \Prefilled{}, and 0.306 in \Coevolving{} (Figure~\ref{fig:solution-diversity}; Table~\ref{tab:solution-diversity}). Three contrast estimates indicated greater concentration, but none differed detectably from its condition-label randomization distribution (all $p\geq.063$).

\begin{table}[t]
\centering
\caption{Exploratory semantic dispersion of proposed packages. Lower leave-one-out cosine distance means greater within-condition concentration. Contrast $p$-values come from 49{,}999 condition-label randomizations.}
\label{tab:solution-diversity}
\small
\begin{tabular*}{\columnwidth}{@{\extracolsep{\fill}}lrr@{}}
\toprule
Condition & $n$ & Mean (SD) distance \\
\midrule
\Reader{} & 181 & 0.340 (0.131) \\
\Chatbot{} & 260 & 0.313 (0.148) \\
\Prefilled{} & 204 & 0.331 (0.163) \\
\Coevolving{} & 151 & 0.306 (0.164) \\
\midrule
Contrast & $\Delta$ & $p$ \\
\midrule
Any AI $-$ \Reader{} & $-0.023$ & .073 \\
\Chatbot{} $-$ \Reader{} & $-0.027$ & .063 \\
Scaffolded $-$ \Chatbot{} & 0.006 & .635 \\
\Coevolving{} $-$ \Prefilled{} & $-0.024$ & .131 \\
\bottomrule
\end{tabular*}
\end{table}

\begin{table}[t]
\centering
\caption{Pairwise comparisons of whether participants interacted with the AI assistant at least once. Odds ratios favor the first-named condition; $p$-values use Holm correction across the three comparisons.}
\label{tab:assistant-use-comparisons}
\small
\begin{tabular*}{\columnwidth}{@{\extracolsep{\fill}}lrrr@{}}
\toprule
Comparison & Odds ratio & 95\% CI & Adjusted $p$ \\
\midrule
\Chatbot{} vs. \Prefilled{} & 2.70 & [1.73, 4.23] & $<.001$ \\
\Chatbot{} vs. \Coevolving{} & 0.46 & [0.22, 0.88] & .014 \\
\Prefilled{} vs. \Coevolving{} & 0.17 & [0.08, 0.32] & $<.001$ \\
\bottomrule
\end{tabular*}
\end{table}